\documentclass[journal,10pt]{IEEEtran}

\usepackage{outline}
\usepackage[utf8]{inputenc} 
\usepackage[T1]{fontenc}    
\usepackage{hyperref}       
\usepackage{url}            
\usepackage{booktabs}       
\usepackage{amsfonts}       
\usepackage{nicefrac}       
\usepackage{pmgraph}
\usepackage[normalem]{ulem}
\usepackage[utf8]{inputenc}
\usepackage{amsbsy,amsthm,amsmath,amssymb,stmaryrd,pict2e,picture}
\hypersetup{colorlinks,allcolors=black}
\usepackage{adjustbox}

\usepackage{graphicx}
\usepackage{times}
\usepackage{xcolor}
\usepackage{xspace}
\usepackage{bm} 
\usepackage{caption}
\usepackage{subcaption}
\usepackage{epstopdf}

\usepackage{enumitem}
\usepackage{graphicx}
\usepackage{epsfig}
\usepackage{tikz}
\usepackage{algpseudocode}
\usepackage{algorithm}
\usepackage{mathrsfs}
\usepackage{setspace}
\usepackage[backend=biber,maxbibnames=10,giveninits=true,
doi=true,isbn=false,url=false,eprint=false,
style=numeric-comp,sorting=none
]{biblatex}

\newtheorem{theorem}{\bf{Theorem}}

\long\def\comment#1{}
\newcommand{\fref}[1] {Fig.~\ref{#1}\xspace}
\newcommand{\tref}[1]{Table~\ref{#1}}

\newcommand{\xtrue}{\xmath{\x_{\mathrm{true}}}}

\DeclareMathOperator*{\argminop}{arg\,min}
\DeclareMathOperator*{\argmaxop}{arg\,max}
\newcommand{\argmin}[1]{\argminop_{#1}} 
\newcommand{\argmax}[1]{\argmaxop_{#1}}
\newcommand{\diag}{\operatorname{diag}}

\newcommand{\xmath}[1] {\ensuremath{#1}\xspace}
\newcommand{\blmath}[1] {\xmath{\bm{#1}}}
\newcommand{\A}{\blmath{A}}

\newcommand{\I}{\blmath{I}}

\newcommand{\x}{\blmath{x}}
\newcommand{\s}{\blmath{s}}
\newcommand{\y}{\blmath{y}}
\newcommand{\z}{\blmath{z}}

\newcommand{\ie}{\text{i.e.}}
\newcommand{\eg}{\text{e.g.}}

\newcommand{\gx}{\xmath{\blmath{g}(\x)}}

\newcommand{\bi} {\xmath{b_i}}

\newcommand{\yi} {\xmath{y_i}}

\newcommand{\defequ} {\triangleq}

\newcommand{\reals} {\xmath{\mathbb{R}}}
\newcommand{\complex} {\xmath{\mathbb{C}}}

\newcommand{\baih} {\xmath{\ba_i'}}

\newcommand{\xk} {\xmath{\blmath{x}_k}}
\newcommand{\xkk} {\xmath{\blmath{x}_{k+1}}}

\newcommand{\ukk} {\xmath{\blmath{u}_{k+1}}}

\renewcommand{\u} {\blmath{u}}
\renewcommand{\v} {\blmath{v}}

\newcommand{\ba}{\boldsymbol{a}}

\newcommand{\br}{\blmath{r}} 

\newcommand{\bw}{\boldsymbol{w}}

\newcommand{\bleta}{\boldsymbol{\eta}}
\newcommand{\bgamma}{\boldsymbol{\gamma}}

\newcommand{\bsigma}{\boldsymbol{\sigma}}
\newcommand{\btheta}{\xmath{\bm{\theta}}}

\def\b{{\blmath b}} 

\newcommand{\paren}[1]{\left( #1 \right)}
\newcommand{\bracket}[1]{\left[ #1 \right]}
\long\def\red#1{\bgroup\color{red}#1\egroup}
\long\def\purple#1{\bgroup\color{purple}#1\egroup}

\newcommand{\todo}[1]{\textbf{\color{red}Todo: #1}}
\newcommand{\cT}{\mathcal{T}}

\definecolor{mich-blue-high}{HTML}{0027CC}
\long\def\blue#1{\bgroup\color{mich-blue-high}#1\egroup}
\long\def\red#1{\bgroup\color{red}#1\egroup}

\makeatletter
\newcommand{\pictslash}[2]{%
  \vcenter{%
    \sbox0{$\m@th#1\varobslash$}\dimen0=.55\wd0
    \hbox to\wd 0{%
      \hfil\pictslash@aux#2\hfil
    }%
  }%
}
\newcommand{\pictslash@aux}[2]{%
    \begin{picture}(\dimen0,\dimen0)
    \roundcap
    \put(0,#1\dimen0){\line(1,#2){\dimen0}}
    \end{picture}%
}
\makeatother

\makeatletter
\newcommand*{\rom}[1]{\expandafter\@slowromancap\romannumeral #1@}
\makeatother

\newcommand{\ymax}{y_{\max}}
\newcommand{\barb}{\xmath{\bar{\b}}}
\newcommand{\gpg}{\xmath{g_{\mathrm{PG}}}}

\def\df{{g}} 
\def\reg{{h}}

\title{Poisson-Gaussian Holographic 
Phase Retrieval \\ 
with Score-based Image Prior}

\author{%
  Zongyu Li$^{*}$, Jason Hu$^*$,
  Xiaojian Xu, Liyue Shen, and Jeffrey A. Fessler
  \\
  EECS Department\\
  University of Michigan \\
  \texttt{zonyul,jashu,xjxu,liyues,fessler@umich.edu}
  \thanks{Z. Li and J. Hu contribute equally to this work. } 
}

\begin{document}

\maketitle

\begin{abstract}
Phase retrieval (PR) is a crucial problem 
in many imaging applications. This study focuses on resolving the holographic phase retrieval problem in situations where the measurements are affected by a combination of Poisson and Gaussian noise, 
which commonly occurs in optical imaging systems.
To address this problem, we propose a new algorithm called ``AWFS" that uses the accelerated Wirtinger flow (AWF) with a score function as generative prior. Specifically, we formulate the PR problem as an optimization problem that incorporates both data fidelity and regularization terms. We calculate the gradient of the log-likelihood function for PR and determine its corresponding Lipschitz constant.
Additionally, we introduce a generative prior in our regularization framework by using score matching 
to capture information about the gradient of image prior distributions.
We provide theoretical analysis that establishes 
a critical-point convergence guarantee
for the proposed algorithm.
The results of our simulation experiments on three different datasets show the following:
1) By using the PG likelihood model, the proposed algorithm improves reconstruction compared to algorithms based solely on Gaussian or Poisson likelihood.
2) The proposed score-based image prior method, 
performs better than the method based on denoising diffusion probabilistic model (DDPM),
as well as plug-and-play alternating direction method of multipliers (PnP-ADMM)
and regularization by denoising (RED).

\comment{
\todo{
\begin{enumerate}
\item Show that Poisson+Gaussian phase retrieval method
is superior than Poisson/Gaussian methods, using unregularized gradient descent (Zongyu).
\item Compare denoisers, such as TV, BM3D, DNCNN, and score-based models, algorithms can be gradient descent or
PnP-ADMM (Jason, Xiaojian).
\item Prove the convergence of gradient descent method with score-based denoiser (Jason, Xiaojian).
\item System matrix: oversampled Fourier with reference image (Zongyu).
\item Dataset: SET11, VIRUS, COIL100.
(Xiaojian: \url{https://drive.google.com/drive/folders/1_bBxFAfBP64JhLyD6c7FTQ0_4xX56e4d?usp=sharing})
\end{enumerate}
}}
\end{abstract}

\section{Introduction}
\label{sec:intro}

Poisson-Gaussian phase retrieval (PR)
is a nonlinear inverse problem,
where the goal is to recover a signal from the (square of) magnitude-only measurements
that are corrupted by 
both Poisson and Gaussian noise \cite{jaganathan:15:pra}. 
This problem is crucial in numerous applications across various fields
such as astronomy \cite{dainty:87:pra}, 
X-ray crystallography \cite{millane:90:pri}, 
optical imaging \cite{shectman:15:prw},
Fourier ptychography
\cite{bian:16:fpr,zhang:17:fpm,xu:18:awf,tian:19:fpr}
and coherent diffractive imaging (CDI)
\cite{latychevskaia:18:ipr}.
In CDI, a coherent beam source illuminates 
a sample of interest and a reference.
When the beam hits the sample,
it generates secondary electromagnetic waves that propagate until they reach a detector.
By measuring the photon flux,
the detector can capture and record a diffraction pattern.
This pattern is roughly proportional to the square of Fourier transform magnitude
of electric field associated with the illuminated objects
\cite{barmherzig:19:drd, barmherzig:19:hpr}.
Recovering the structure of the sample
from its diffraction pattern is an non-linear inverse problem known as holographic PR.
To solve this problem, the maximum a posterior (MAP) estimate can be conducted with the following form:
\begin{equation}\label{eq:phase,cost}
\hat{\x} = 
\argmax{\x \in \reals^N} p(\x|\y,\A,\br)
=
\argmin{\x \in \reals^N}{\df(\x;\A,\y, \br) + \reg(\x)},  
\end{equation}
where \x denotes the image to recover, 
\y is the measurement collected,
$\A \in \complex^{M \times N}$ denotes 
the corresponding system matrix
in holographic PR 
where $M$ denotes the number of measurements and
$N$ denotes the dimension of \x. 
The known reference image
\br
provides additional information to reduce the ambiguity of $\hat{\x}$;
using an extended reference is a common technique in Holographic CDI
\cite{saliba:12:fth, guizar-sicairos:07:hwe}.
Following Bayes' rule, 
we denote 
$\df(\x) = -\log p(\y,\A, \br|\x)$ 
and $\reg(\x) = -\log p(\x)$ 
as the data-fidelity term
and the regularization term, respectively.

\begin{figure*}[t]
    \centering
    \includegraphics[width=\linewidth]{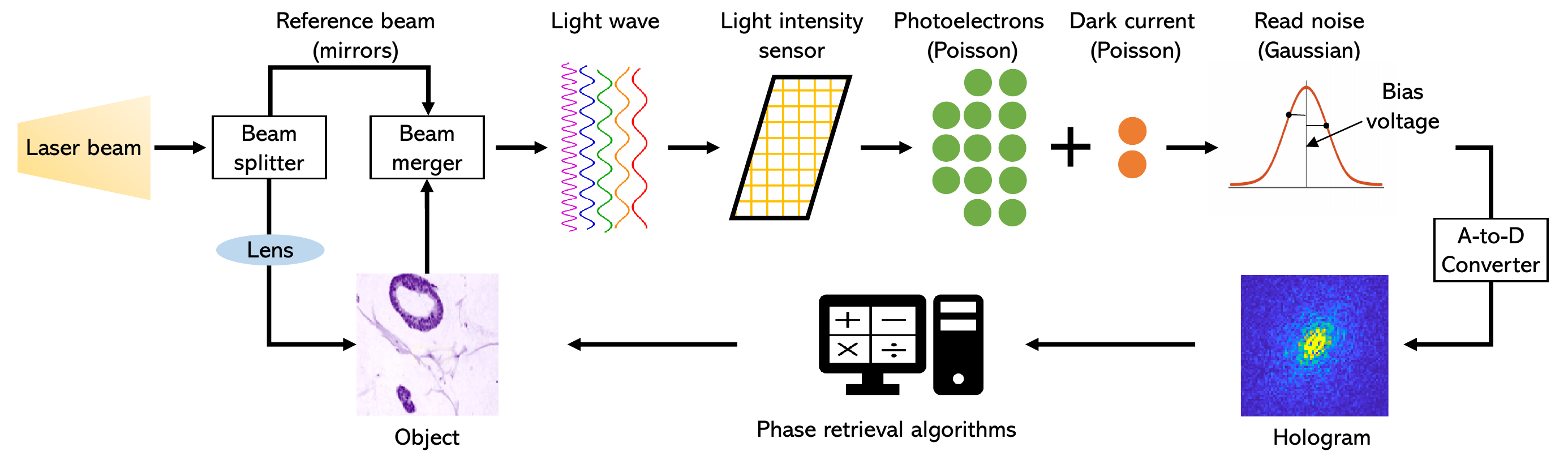}
    \caption{~\emph{Illustration of Poisson and Gaussian noise statistics in holographic phase retrieval.}}
    \label{fig:pgphase-illustration}
\end{figure*}

In practical scenarios,
the measurements \y are contaminated 
by both Poisson and Gaussian (PG) noise.
The Poisson distribution
is due to the
photon counting and dark current \cite{wang:19:tpc}.
The Gaussian statistics stem from the readout structures 
(\eg, analog-to-digital converter (ADC)) 
of common cameras.
\fref{fig:pgphase-illustration}
illustrates the PG mixed noise statistics 
in the holographic PR.
Because the PG likelihood is complicated (see \eqref{eq:pois,gau,likelihood}), 
most previous works
\cite{candes:13:pea, candes:13:prv, shectman:15:prw, jiang:16:wfm, cai:16:oro, soltanolkotabi:19:ssr,
qiu:16:ppr, cai:16:oro, gerchberg:72:paf, netrapalli:15:pru, waldspurger:18:prw,
gao:17:pru,li:16:ogd,liang:18:prv,yang:19:pcd, sun:16:aga, candes:15:prv, wu:19:orb,
thibault:12:mlr,goy:18:lpc,xu:18:awf,barmherzig:20:lph,vazquez:21:qpr,
lawrence:20:prw,
gnanasambandam:20:ici, 
li:22:ppr, fatima:21:panp, chen:17:srq, bian:16:fpr, zhang:17:ana, chang:18:tvb}
approximate the Poisson noise statistics
by the central limit theorem
and work with a substitute Gaussian log-likelihood estimate problem;
or use the Poisson maximum likelihood model
but simply disregards Gaussian readout noise.
Besides, other more complicated approximation methods have also been proposed,
such as shifted Poisson model
\cite{snyder:93:irf},
the unbiased inverse transformation
of a generalized Anscombe transform
\cite{Makitalo:13:oio,tian:19:fpr}
and the majorize-minimize algorithm \cite{fatima:22:pam}.
However, these approximate methods 
can lead to a suboptimal solution after optimization 
that results in a lower-quality reconstruction.

Apart from the likelihood modeling,
the regularizer $h(\x)$ provides prior information about 
underlying object characteristics that 
may aid in resolving ill-posed inverse problems.
Beyond simple choices of $h(\x)$
such as total variation (TV)
or the L1-norm of coefficients of wavelet transform
\cite{daubechies:92:tlo},
deep learning (DL)-integrated algorithms 
for solving inverse problems in computational imaging
have been reported to 
be the state-of-the-art \cite{ongie:20:dlt}.
The trained networks can be used as an object prior
for regularizing the reconstructed image to remain on a learned manifold \cite{bora:17:csu}.
Incorporating a trained denoising network 
as a regularizer $\reg(\cdot)$
led to methods such as
plug-and-play (PnP) 
\cite{chan:17:pap, zhang:22:pap, kamilov2022pnp},
and regularization by denoising (RED) \cite{romano:17:tle}.
In contrast to training a denoiser using clean images,
there is growing popularity of self-supervised image denoising approaches
that do not require clean data as the training target \cite{lehtinen:18:nli, batson:19:nbd, wang:22:bss}.

In addition to training a denoiser as regularizer,
generative model-based priors are also proposed \cite{asim:20:igm, wei:22:duw}.
Recently, diffusion models have gained significant attention
for image generation
\cite{song:19:gmb, ho:20:ddp, dhariwal:21:dmb, song:21:sbg}.
These probabilistic image generation models start with a clean image and 
gradually increase the level of noise 
added to the image,
resulting in white Gaussian noise.
Then in the reverse process, a neural network is trained
to learn the noise in each step to generate or sample a clean image
as in the original data distribution. 
The score-based diffusion models 
estimate the gradients of data distribution
and can be used as plug-and-play priors for the inverse problem solving \cite{graikos:22:dma}
such as image delurring, 
MRI and CT reconstruction
\cite{lee:22:pdo, jalal:21:rcs, chung:22:sbd, cui:22:sss, long:09:a3f, song:22:sip}. 
However, the realm of using score-based models to perform phase retrieval is relatively unexplored;
previous relevant works \cite{shoushtari:22:dolph, graikos:22:dma} 
applied denoising diffusion probabilistic modeling (DDPM) to PR
but with less realistic system models
and under solely Gaussian or Poisson noise statistics.

In summary, our contribution is in three-fold:
\vspace{-.8em}
\begin{itemize}[leftmargin=*]
\item 
We present a new algorithm known as accelerated Wirtinger flow with score-based image prior
(\ie, $\nabla \reg(\x)$ in \eqref{eq:phase,cost}) to address the challenge of Holographic phase retrieval (PR) problem in the presence of Poisson and Gaussian (PG) noise statistics.
\item
Theoretically, we derive a Lipschitz constant for the Holographic PR's PG log-likelihood and subsequently demonstrate the critical points convergence guarantee of our proposed algorithm.
\item 
Simulation experiments demonstrate that:
1) Algorithms with the PG likelihood model yield superior reconstructions in comparison to those relying solely on either the Poisson or Gaussian likelihood models. 
2) With the proposed score-based prior as regularization,
the proposed approach generates higher quality reconstructions
and is more robust to variation of noise levels without any parameter tuning, 
when compared to alternative state-of-the-art methods.
\end{itemize}


\section{Background}

This section reviews
holographical phase retrieval
and machine learning-based regularizers for inverse problem solving.

\subsection{Phase Retrieval (PR)}

\textbf{Gaussian PR.}
By assuming the elements of \y
follow independent Gaussian distributions
\(\y \sim \mathcal{N}(|\A\x|^2 + \barb, \sigma^2 \I)\),
the data fidelity term $g(\x)$ in \eqref{eq:phase,cost} 
becomes
$\df_{\mathrm{Gau}}(\x) \defequ \|\y - \barb - |\A\x|^2\|_2^2$.
To solve the corresponding MAP optimization problem,
a popular method is Wirtinger flow (WF)
\cite{candes:15:prv,jiang:16:wfm, cai:16:oro, soltanolkotabi:19:ssr}
using the Wirtinger gradient:
$
\nabla \df_{\mathrm{Gau}}(\x) = 4\A'\diag\{|\A\x|^2 - \y + \barb\}\A\x.    
$
To determine an appropriate step size
for the Wirtinger gradient,
one can utilize the Lipschitz constant, 
or consider using methods 
such as empirical trial and error, backtracking line search or observed Fisher information \cite{li:22:ppr}.
To further accelerate the WF, 
one can use Nesterov's momentum methods 
\cite{nesterov:05:smo}
or optimized gradient methods \cite{kim:16:ofo},
leading to the accelerated Wirtinge flow (AWF)
\cite{xu:18:awf, bostan:18:awf, fabian:20:3pr, gao:22:psr}
that is commonly used in solving PR problems.
Apart from WF, other methods such as matrix-lifting 
\cite{candes:13:pea, candes:13:prv, shectman:15:prw}, 
Gerchberg-Saxton \cite{gerchberg:72:paf},
majorize-minimize \cite{qiu:16:ppr} and 
alternating direction method of multipliers (ADMM) \cite{liang:18:prv}
have also been proposed.

\textbf{Poisson PR.}
The Poisson ML model assumes
\(\y \sim \mathrm{Poisson}(|\A\x|^2 + \barb)\), 
so that $g(\x)$ in \eqref{eq:phase,cost} has the form:
$
\df_{\mathrm{Pois}}(\x) \defequ \blmath{1}'(|\A\x|^2 + \barb) 
- \y'\log(|\A\x|^2 + \barb).
$
Similar to the Gaussian case, one can also apply 
WF \cite{li:21:ppr} with
$\nabla \df_{\mathrm{Pois}}(\x) = 2 \A\x \odot \paren{\blmath{1} - \y \varoslash (|\A\x|^2 + \barb)},$
where $\odot$ and $\varoslash$ denote element-wise multiplication
and division, respectively.

Algorithm~\ref{alg:wf} summarizes the WF approach for PR. The Gaussian and Poisson models
are both suboptimal
for practical scenario where
the measurements are corrupted with Poisson plus Gaussian noise.
Furthermore, many previous works
\cite{candes:13:pea,waldspurger:18:prw,gao:17:pru,li:16:ogd,yang:19:pcd, sun:16:aga, candes:13:prv, jiang:16:wfm,
cai:16:oro, soltanolkotabi:19:ssr, zhang:17:ana, fatima:22:pam, shoushtari:22:dolph, wu:19:orb, li:21:ppr}
modeled the system matrix \A (in \eqref{eq:phase,cost})
as i.i.d. random Gaussian
or randomly masked Fourier transform;
these assumptions simplify the PR problem  
and lead to elegant mathematical derivations 
(\eg, spectral initialization \cite{candes:15:prv, luo:19:osi}),
but they are less related to optical imaging systems used in practical PR \cite{zhuang:22:ppr}.
Practical PR involves
canonical Fourier transform-based system matrices such as
Fresnel PR \cite{wang:20:piw, zhang:21:pad},
holographic PR \cite{barmherzig:22:tph},
ptychographic PR \cite{thibault:08:hrs, marchesini:13:apf}
and Fraunhofer PR \cite{siu:03:uxp, zhuang:22:ppr}.
Some previous works also
remove the square of the Fourier transform magnitude (see \eqref{eq:pois,gau,stat})
\cite{sun:16:aga, zhang:17:ana, shoushtari:22:dolph, wu:19:orb}.
However,
this square of magnitude indicates the 
amount of wavelength-weighted power emitted by a light source per unit area, 
so its removal reduces the practicality.

\begin{algorithm}[t]
\caption{Phase retrieval via Wirtinger Flow}
\label{alg:wf}
\begin{algorithmic}
\Require Measurement $\y$, system matrix $\A$, 
initialization of image $\x_0$, regularizer $h(\cdot)$.
\For{$k=1:K$}
\If{Gaussian noise model is used}
\State 
Compute $\nabla \df_{\mathrm{Gau}}(\xk) = 4\A'\diag\{|\A\xk|^2 - \y + \barb\}\A\xk.$
\ElsIf{Poisson noise model is used}
\State 
Compute $\nabla \df_{\mathrm{Pois}}(\xk) = 2 \A\xk \odot (\blmath{1} - \y \varoslash (|\A\xk|^2 + \barb)).$
\EndIf
\State Compute gradient of the regularizer $\nabla h(\xk)$.
\State Compute step size $\mu_k$. 
\State Set $\xkk = \xk - \mu_k \paren{\nabla g(\xk) + \nabla h(\xk)}$.
\EndFor
\end{algorithmic}
Return $\x_{K}$
\end{algorithm}

\subsection{Deep Learning-based Regularizers}

\textbf{PnP and RED.}
Both PnP and RED are widely adopted in a variety of inverse problems~\cite{Sreehari.etal2016,Ahmad.etal2020,Zhang.etal2017a,Zhang.etal2019}.
By implicitly representing the prior $\reg(\cdot)$ in~\eqref{eq:phase,cost} by an image denoiser,
plug-and-play (PnP) methods~\cite{Venkatakrishnan.etal2013, Kamilov.etal2017} were proposed
to allow the integration of the physical measurement models and powerful DL-based denoisers as image priors \cite{kamilov2022pnp}.
The model-agnostic nature of the denoiser
allows PnP methods to be applied to multiple imaging problems 
using a single DL denoiser
simply by changing the imaging model.
Regularization by denoising (RED)~\cite{Romano.etal2017, Reehorst.Schniter2019}
is an algorithm closely related to PnP
that uses denoising engine in defining the regularization of the inverse problems.
\textbf{Score Function and Diffusion Models.}
Let $p_{\btheta}(\x)$ denote a model for the prior distribution of the latent image \x;
the score function is then defined as%
\footnote{This definition differs from the score function in statistics
where the gradient is taken w.r.t. \btheta
of $\log p_{\btheta}(\x)$.}
$s_{\btheta}(\x) = \nabla_{\x} \log p_{\btheta}(\x)$.
Consider a sequence of positive noise scales
(for white Gaussian $\mathcal{N}(0, \sigma_k^2)$): $\sigma_1 < \sigma_2 < \cdots < \sigma_K$, 
with $\sigma_1$ being small enough so that noise of this level does not visibly affect the image, 
and $\sigma_K$ depending on the application.
Score matching can be used to train a noise conditional score network (NCSN)
\cite{vincent:11:acb,song:19:gmb}
as follows: 
\begin{align}\label{eq:score,function,theta}
     &\hat{\btheta} = 
     \argmin{\btheta}{\sum_{k=1}^K \mathbb{E}_{\x, \tilde{\x}}
     \bracket{\paren{ s_{\btheta}(\x, \sigma_k) - \frac{\x-\tilde{\x}}{\sigma_k^2}}^2}},
     \nonumber \\
     & \mathrm{where} \quad \x \sim p(\x), \quad \tilde{\x} \sim \x+\mathcal{N}(0, \sigma_k^2 \I).
\end{align}
With enough data, the neural network $s_{\btheta}
(\x, \sigma)$ is expected to learn the distribution 
$p_\sigma(\x) = \int p(\x) p_\sigma (\y|\x)d\x$ 
where $p_\sigma (\y|\x) = \mathcal{N} (\x, \sigma^2 \I)$
and $\y$ is defined in \eqref{eq:phase,cost}. 
To sample from the prior, the method of Langevin dynamics is frequently used \cite{song:19:gmb}.

To leverage diffusion models for solving inverse problems, 
previous methods generally recast the reconstruction problem 
as a conditional generation or sampling problem \cite{shoushtari:22:dolph, graikos:22:dma, song:21:sbg, chung:22:idm, song:22:sip, chung:23:dps}.
This involves relying on the capacity of diffusion models to produce high-quality images while complying with data-fidelity constraints.
However, in applications where the data collection is costly, 
\ie, with a limited amount of training data, 
it is often challenging to obtain a diffusion model 
that can generate high-quality images 
even in an unconditional way. 
With such a scenario, we found that the score function learned during training diffusion models can serve as 
an effective image prior 
(as demonstrated in Section~\ref{sec:experiment}), 
which can capture certain data characteristics 
when trained for the denoising prediction in the reverse process of the diffusion model. Similar to previous works \cite{graikos:22:dma} that uses the score function as a PnP prior, here we also incorporate the score function 
as a regularization in the optimization objective for solving the PR problem. We think this is a more efficient scheme for incorporating diffusion priors especially for applications with a limited amount of training data that is a very common situation in the optical imaging sector.

\comment{run 
\begin{equation}
    x_i^m = x_i^{m-1} + \epsilon_i s_\theta (x_i^{m-1}, \sigma_i) + \sqrt{2\epsilon_i} z_i^m,
\end{equation}
where $m$ ranges from 1 to $M$, and $z_i^m$ is a standard Gaussian.
}
\section{Methods}
\label{sec:method}

\subsection{Wirtinger Flow (WF)}
\label{subsec:wf}
Based on the physical model as demonstrated 
in \fref{fig:pgphase-illustration},
we model the system matrix \A
by the (oversampled and scaled) discrete Fourier transform
applied to a concatenation of the sample \x, 
a blank image 
(representing the holographic separation condition \cite{lawrence:20:prw})
and a known reference image \br,
so that \y follows the following distribution:
\begin{align}
\label{eq:pois,gau,stat}
&\y \sim \mathcal{N}(
\mathrm{Poisson}
\paren{
|
\A(\x)
|^2 + \barb
}
, \sigma^2 \I),
\nonumber \\
&\A(\x) \defequ \alpha \mathcal{F}\{[\x, \blmath{0}, \br]\}.
\end{align}
Here \barb denotes the mean of background measurements,
$\sigma^2$ denotes the variance of Gaussian noise,
and $\alpha$ denotes a scaling factor 
(being quantum efficiency, conversion gain, etc)
after applying the 
Fourier transform.
Plugging the 
negative log-likelihood of \eqref{eq:pois,gau,stat} 
into \eqref{eq:phase,cost} leads to
\begin{align}
\label{eq:pois,gau,likelihood}
&\df_{\mathrm{PG}}(\x) = \sum_{i=1}^M \df_i(\x),
\\
&
\resizebox{\linewidth}{!}{
$\df_i(\x) \defequ -\log\paren{
\sum_{n=0}^{\infty}
\frac{e^{-\paren{|\baih \x|^2 + \bar{b}_i}}\cdot \paren{|\baih \x|^2 + \bar{b}_i}^n}{n!}
\cdot \frac{e^{-\paren{\frac{(\yi -n)^2}{\sqrt{2}\sigma}}}}{\sqrt{2\pi \sigma^2}}
}
.$}
\nonumber
\end{align}
Here $M$ denotes the length of $\y$, $\baih$ denotes the 
$i$th row of $\A$ (since $\A$ is linear).
We opt to use WF for estimating $\x$ 
because it is commonly used in practice due to
its simplicity and efficiency \cite{candes:15:prv}.
The WF algorithm is based on the gradient of \eqref{eq:pois,gau,likelihood}:
\begin{align}
\label{eq:pois,gau,grad}
&\nabla \df_{\mathrm{PG}}(\x) 
= 2 \A' \diag\{ \phi_i(|\baih \x|^2 + \bi; \yi)
\} 
\A \x,
\\
&\phi(u; v) \defequ
1 - \frac{s(u, v - 1)}{s(u, v)},
\quad 
s(a, b) \defequ 
\sum_{n=0}^{\infty}
\frac{a^n}{n!} e^{-\paren{\frac{b-n}{\sqrt{2}\sigma}}^2}
.
\nonumber
\end{align}

\textbf{Lemma 1.}\label{lemma,lips}
The function $\phi(u)$ is 
Lipschitz differentiable 
and the Lipschitz constant for $\dot\phi(u)$ is:
\begin{align}\label{eq:lemma,lips}
&\max\{|\ddot{\phi}(u)|\} \defequ \mu = \paren{1 - e^{-\frac{1}{\sigma^2}}}
e^{\frac{2 y_{\max} - 1}{\sigma^2}
},
\nonumber \\
&\mathrm{where}
\quad 
y_{\max} = \underset{i \in \{1,\ldots,M\}}{\max} 
\{\yi\}
.
\end{align}
The proof of Lemma 1 is given in \cite{chouzenoux:15:aca}.

\begin{theorem}
\label{theorem,wf,lip}
Assume $|x_j|$ is bounded above by 
$C$ for each $j$, a Lipschitz constant of $\nabla \gpg(\x)$ is 
\begin{align}
\label{eq:pois,gau,lips} 
\mathcal{L}(\nabla 
&\df_{\mathrm{PG}}) \defequ
\, 
4C^2 \|\A\|_2^2\,
\|\A\|_{\infty}^2\,
\paren{1 - e^{-\frac{1}{\sigma^2}}}
\,e^{\frac{2 y_{\max} -1}{\sigma^2}
}
\nonumber \\
&+2\|\A\|_2^2 \Big|1 - C^2\, \|\A\|_{\infty}^2\,
\paren{1 - e^{-\frac{1}{\sigma^2}}}
\,e^{\frac{2 y_{\max} -1}{\sigma^2}
}
\Big|
,    
\end{align}
where $\ymax$ is $\max_i \{\|\yi|\}, i=1,\ldots,M$.
\end{theorem}

\textbf{Proof:}
Let $\gpg(\x)$ denote a function that maps 
a vector $\x \in \reals^N$ to a scalar;
it is the sum of each $g_i(\x) \defequ \phi_i(|\baih \x|^2 + \bi; \yi)$
over $i=1,\ldots, M$.
Let \gx denote a function
that maps 
a vector $\x \in \reals^N$ 
to the measurement space 
$\y \in \reals^M$;
it is the concatenation of each $g_i(\x)$.
So $\nabla \gpg(\x) \in \reals^N$
, $\nabla^2 \gpg(\x) \in \reals^{N\times N}$,
and $\nabla \gx \in \reals^{M\times N}$.

By the chain rule,
the Hessian of \gpg 
is
\begin{equation}\label{eq:gpg,hessian}
\nabla^2 \gpg(\x) = 
2\A'\paren{
\diag\{\A\x\} \nabla \gx
+ \diag\{ \gx \} \A
}
.\end{equation}
Assume $|x_j|$ is bounded above by 
$C$
for each $j$.
Then it follows that
$\|\diag\{\A\x\}\|_2 \le C\|\A\|_{\infty}$
by the construction of matrix-vector multiplication,
leading to a Lipschitz constant for $\nabla \gpg(\x)$: 
\begin{equation}\label{eq:gpg,lips,1}
\resizebox{\linewidth}{!}{$\mathcal{L}(\nabla \gpg) = 
2C\|\A\|_2 \, \|\A\|_{\infty} \, \|\nabla \gx\|_2 
+ 2\|\A\|_2^2 \, \|\diag\{\gx\}\|_2
.$}
\end{equation}
Here $\mathcal{L}(\nabla \gpg)$ denotes a Lipschitz constant for $\nabla \gpg$,
not necessarily the best one.
To compute
$\|\nabla \gx \|_2$,
we substitute the Lipschitz constant of $\dot\phi(u)$
into \eqref{eq:pois,gau,grad} and apply Lemma 1,
leading to
\begin{equation}\label{eq:grad,g}
\|\nabla \gx\|_2 
\le 
2C \|\A\|_2 \|\A\|_{\infty}
\paren{1 - e^{-\frac{1}{\sigma^2}}}
e^{\frac{2 y_{\max} -1}{\sigma^2}}
.
\end{equation}
To compute $\|\diag\{\gx\}\|_2$,
let
\begin{equation}\label{eq:t,bound}
t \in 
[b, \max_i\{|\baih \x|^2\} + b] 
\subseteq
\cT \defequ
[b, C^2 \|\A\|_{\infty}^2 + b].
\end{equation}
From the fact that 
\(\dot{\phi}(t) \le 1\)
by its construction,
one can derive that
\begin{align}\label{eq:diag,gi,norm}
\|\diag\{ \gx \}\|_2
= \| \gx \|_{\infty}
&\leq 
\max_{t \in \cT} \{|\dot{\phi}(t)|\}
\\
&\le 
\Big|1 - C^2 \|\A\|_{\infty}^2
\max\{|\ddot{\phi}(t)|\}\Big|
.\nonumber
\end{align}

Combining \eqref{eq:gpg,lips,1}, \eqref{eq:grad,g}
and \eqref{eq:diag,gi,norm} completes the proof
of Theorem 1. \qed 

Theorem~\ref{theorem,wf,lip}
is an extension of \cite{chouzenoux:15:aca}
that considers a linear transformation model
to a non-linear transformation 
($\A\x \rightarrow |\A\x|^2$)
and with a different system matrix \A.

However, due to the infinite sum in Poisson-Gaussian WF
\eqref{eq:pois,gau,likelihood},
we approximate the $s(a, b)$ following \cite{chouzenoux:15:aca}:
\begin{equation}
\label{eq:truncate,sum}
s(a, b)
\approx 
\sum_{n=0}^{n^{+}}
\frac{a^n}{n!}e^{-\paren{\frac{b-n}{\sqrt{2}\sigma}}^2},
\quad 
n^{+} = \lceil n^* + \delta \sigma \rceil,
\end{equation}
with $n^*$ given by 
\begin{align}
\label{eq:lambert}
n^* &= \sigma \mathcal{W}\paren{\frac{a}{\sigma^2}e^{b/\sigma^2}}
\nonumber \\
&\approx
\sigma \paren{
\frac{b}{\sigma^2}
\log
\paren{
\frac{a}{\sigma^2}
}
- \log\paren{
\frac{b}{\sigma^2}
\log
\paren{
\frac{a}{\sigma^2}
}
}
}
\nonumber
\\&
= 
\frac{b}{\sigma}
\log
\paren{
\frac{a}{\sigma^2}
}
- \sigma 
\log\paren{
\frac{b}{\sigma^2}
\log
\paren{
\frac{a}{\sigma^2}
}}
,
\end{align}
where $\mathcal{W}(\cdot)$ denotes the Lambert function.
The accuracy of this approximation 
is controlled by $\delta$.
Reference \cite{chouzenoux:15:aca} did a comprehensive analysis 
on the maximum error value of the truncated sum \eqref{eq:truncate,sum} and found the bound was quite precise.

\subsection{Accelerated Wirtinger Flow
with Score-based Image Prior}

For the acceleration scheme in the WF algorithm,
we followed the implementation of \cite{li:15:apg}
as its convergence guarantee was proved.
Under the assumption that the true score function can be learned properly,
when we have a trained score function
$s_{\btheta} (\x, \bsigma)$ by applying \eqref{eq:score,function,theta},
the gradient descent algorithm 
for MAP estimation \eqref{eq:phase,cost} has the form:
$\x_{t+1} = \x_t - \mu (\nabla \df(\x_t) + s_{\btheta} (\x_t, \sigma_k))$.
Algorithm~\ref{alg:pg_score}
summarizes our proposed AWFS algorithm
(the vanilla version without acceleration is given in Appendix~\ref{appendix, alg}).
In a similar fashion as Langevin dynamics, we choose $\sigma_k$ to be a descending scale of noise levels. In practice, we generally use each noise level a fixed number of times, with geometrically spaced noise levels between some lower and upper bound. 
The stepsize factor $\epsilon$ in Algorithm~\ref{alg:pg_score}
can be selected empirically,
but we show that the Lipschitz constant of the gradient 
$\nabla \df_{\mathrm{PG}}(\x_t)
+
s_{\btheta} (\x_{t}, \sigma_k)$ 
exists as demonstrated in Theorem~\ref{theorem, score, lip}
(the proof is given in the Appendix~\ref{appendix, p2});
hence with sufficiently small step size $\epsilon$,
the sequence generated by
Algorithm~\ref{alg:pg_score} will converge to 
a critical point of 
the posterior distribution in \eqref{eq:phase,cost}.

$\textbf{Lemma 2.}$
The Fourier transform (and inverse transform) of an absolutely integrable function is continuous.

$\textbf{Lemma 3.}$
Suppose a sequence of functions
$f_i: \mathbb{R} \rightarrow \mathbb{R}$
converges in the $L^1$ to some function $f$, 
and that each $f_i$ is absolutely integrable.
Then $f$ is also absolutely integrable.

\textbf{Proposition 1.}
The derivative of 
$\log (p(x) \circledast \mathcal{N}(0, \sigma^2))$
is bounded on the interval $[-C,C]$.

A brief proof of Proposition 1 can be started by using
derivatives of convolution, \ie, 
\begin{align}\label{eq:derivative, px}
    &\frac{d}{dx}(\log (p(x) \circledast \mathcal{N}(0, \sigma^2)))
    \nonumber \\
    &\sim \frac{\mathcal{F}^{-1}(\imath x \mathcal{F}(p(x)) \cdot
    \mathcal{F}(\mathcal{N}(0, \sigma^2)))}{p(x) \circledast \mathcal{N}(0, \sigma^2)}
    \nonumber \\
    &\sim \frac{\mathcal{F}^{-1}(xe^{-x^2} \cdot \mathcal{F}(p(x)))}{p(x) \circledast \mathcal{N}(0, \sigma^2)}
    .
\end{align}
And then use Lemma 2 and 3 and the fact that
a sequence of Gaussian mixture models (GMMs) can be used
to approximate any smooth probability distribution in $L^2$ convergence \cite[pp.~65]{goodfellow:16:dl}. 
The complete proof can be found in the appendix.

\textbf{Lemma 4}:
Suppose 
$f(x,y):\reals^2 \rightarrow \reals$ is an everywhere twice differentiable function.
Then $\frac{\partial^2}{\partial x \partial y} \log f(x,y)$ is bounded
if the following three conditions are met:
\begin{enumerate}
\item $\frac{\partial^2}{\partial x \partial y} f(x,y)$ is bounded.
\item $f$ itself is bounded below by a positive number and also bounded above.
\item $\nabla f$ is bounded.
\end{enumerate}

\textbf{Proposition 2.}
The gradient of $p_\sigma(\x)$ is Lipschitz continuous
on $[-C,C]^N$.

By the definition of Lipschitz continuity, 
it suffices to show the Hessian of
$ p_\sigma(\x) = p(\x) \circledast \mathcal{N}(0, \sigma^2 \I)$ has bounded entries.
By renaming the variables, and redefining $f(x,y) = p(x,y, \cdots)$,
we may consider the boundedness of
$\frac{\partial^2}{\partial x \partial y} \left( \log f(x,y) \circledast \mathcal{N}(0, \sigma^2 \I) \right)$
on $[-C,C]^2$ and use Lemma 4 to remove the log.
The appendix shows the full proof of Proposition 2.

\begin{algorithm}[t]
\caption{Our proposed accelerated WF with score-based image prior.}
\label{alg:pg_score}
\begin{algorithmic}
\Require Measurement \y, system matrix \A, 
momentum factor $\eta_0=1$,
step size factor $\epsilon$,
weighting factor $\bgamma$,
truncation operator $\mathcal{P}_C(\cdot) \rightarrow [0, C]$;
initial image $\x_0$,
initial auxiliary variables $\z_0 = \bw_0 = \v_0 = \x_0$, 
initialize $\sigma_1 > \sigma_2 > \cdots > \sigma_K$.
\For{$k=1:K$}
\For{$t=1:T$}
\State Set step size $\mu = \epsilon \sigma_k^2$.
\State Set $\Delta z_{t,k} = \frac{\eta_{t-1,k}}{\eta_{t,k}}
(\z_{t,k} - \x_{t,k})$.
\State Set $\Delta x_{t,k} = \frac{\eta_{t-1,k}-1}{\eta_{t,k}}
(\x_{t,k} - \x_{t-1,k})$.
\State Set $\bw_{t, k} = 
\mathcal{P}_{C}\paren{\x_{t,k} + \Delta z_{t,k}
+ \Delta x_{t,k}}$.
\State Compute $s_{\btheta} (\x_{t,k}, \sigma_k)$
and $s_{\btheta} (\bw_{t,k}, \sigma_k)$.
\State Set \resizebox{0.81\linewidth}{!}{$\z_{t+1,k}=\bw_{t,k} - \mu 
\paren{\nabla \gpg(\bw_{t,k}) + s_{\btheta} (\bw_{t,k}, \sigma_k)}$.}
\State Set $\v_{t+1,k}=\x_{t,k} - \mu 
\paren{\nabla \gpg(\x_{t,k}) + s_{\btheta} (\x_{t,k}, \sigma_k)}$.
\State Set $\eta_{t+1, k} = \frac{1}{2}\paren{1 + \sqrt{1 + 4 \eta_{t,k}^2}}$.
\State Set $\x_{t+1,k} = 
\mathcal{P}_{C}\paren{
\gamma_{t,k}\z_{t+1,k} 
+ (1-\gamma_{t,k})\v_{t+1,k}}$.
\EndFor
\EndFor
\\ 
Return $\x_{T,K}$.
\end{algorithmic}
\end{algorithm}

\begin{theorem}\label{theorem, score, lip}
For a smooth density function $p(\x)$ that has finite expectation, 
the Lipschitz constant of 
$\nabla \df_{\mathrm{PG}}(\x_{t,k})
+
s_{\btheta} (\x_{t,k}, \sigma_k)$
exists when each element in $\x_{t,k}$ 
satisfy $0<|x_j|<C$ for each j.
Furthermore, if the weighting factor $\bgamma \in \{0,1\}$
is chosen appropriately following \cite{li:15:apg},
\ie,
according to 
whichever higher posterior probability 
between $p(\z|\y,\A,\br)$ and $p(\v|\y,\A,\br)$;
then with sufficiently small $\epsilon$,
the sequence
$\{\x_{t,k}\}$
generated by Algorithm~\ref{alg:pg_score} is bounded,
and any accumulation point of $\{\x_{t,k}\}$
is a critical point of the posterior distribution
$p(\x|\y,\A,\br)$ in \eqref{eq:phase,cost}.
\end{theorem}

\textbf{Proof:} By Proposition 2,
and from the design of Algorithm~\ref{alg:pg_score},
$\x_{t,k}$ and $\bw_{t,k}$ are 
both bounded between 
$[0, C]$ for all $t,k$,
so the Lipschitz constant $\mathcal{L^*}$ 
of $\nabla \df_{\mathrm{PG}}(\cdot)
+
s_{\btheta} (\cdot)$ exists.
With the stepsize $\mu$ satisfying 
$0<\mu<\frac{1}{\mathcal{L^*}}$,
and the weighting factor $\bgamma \in \{0,1\}$
being chosen according to 
whichever higher posterior probability 
between $p(\z|\y,\A,\br)$ and $p(\v|\y,\A,\br)$
(see \cite{li:15:apg}),
then we satisfy all conditions in Theorem 1 of \cite{li:15:apg},
which establishes the critical-point convergence of Algorithm~\ref{alg:pg_score}. \qed

Theorem~\ref{theorem, score, lip} states that $\bgamma \in \{0,1\}$ 
is needed to achieve convergence; 
nevertheless,  
we empirically discovered that setting $\bgamma=0.5$ 
also led to promising performance.
Compared to our score-based image prior approach,
another method is to alternate data-consistency updates with 
the sampling step of a diffusion model based on 
denoising diffusion probabilistic models (DDPM) framework \cite{shoushtari:22:dolph},
as shown in Algorithm~\ref{alg:dolph}.

\begin{algorithm}[ht]
\caption{DOLPH \cite{shoushtari:22:dolph}.}
\label{alg:dolph}
\begin{algorithmic}
\Require Measurement $\y$, 
system matrix $\A$, 
initialization of image $\x_T$, 
pre-trained DDPM model $s_{\btheta}$,
and $T$.
\For{$t=T:1$}
\State Set $\z_t \sim \mathcal{N}(0, \I)$ if $t>1$ and $\z_t =0$ otherwise.
\State Set $\x_{t-1} = \frac{1}{\sqrt{\alpha_t}} \left (\x_t - \frac{1-\alpha_t}{\sqrt{1-\overline{\alpha_t}}} \s_{\theta}(\x_t, t) \right ) + \sigma_t \z_t$.
\State Determine stepsize $\mu_k$.
\State Set $\x_{t-1} = \x_{t-1}-\mu_k \nabla \gpg(\x_{t-1})$.
\EndFor
\end{algorithmic}
Return $\x_1$.
\end{algorithm}

\section{Experiment}
\label{sec:experiment}
\subsection{Experiment Settings}
\textbf{Dataset}.
We tested all algorithms on three dataset:
162 histopathology images related to breast cancer \cite{aksac:19:bad} (train/val/test is 122/20/20);
920 images from CelebA dataset \cite{liu::15:dlf}
(train/val/test is 800/100/20);
and 720 images from a homemade CT-density dataset
(train/val/test is 600/100/20).
The CT-density dataset was generated from 
SPECT/CT images for Yttrium‐90 radionuclide therapy
after applying the CT-to-density calibration curve
\cite{li:22:dad}.
Although the size of training datasets are relatively small compared to typical datasets such as ImageNet or LSUN \cite{ho:20:ddp, song:21:sbg} that have millions of images,
we do not require the score functions to learn image priors strong enough to generate realistic images from white Gaussian noise; rather, it is sufficient for the priors to be able to denoise moderately noisy images.

\textbf{System Model}.
The system matrix is based on discrete Fourier transform
of the concatenation of the true image \x,
a blank image $\blmath{0}$
and a reference image \br
with scaling and oversampling.
We set the scaling factor $\alpha$ 
to be in the range $ [0.02, 0.035]$
so that the average counts per pixel 
range from 6 to 25;
the oversampled ratio is set to 2.
We set $\br$ to be a binary random image similar to what was used in \cite{lawrence:20:prw}.
The standard deviation of the Gaussian read noise 
added to the measurements \y was set as
$\sigma \in [0.5, 1.5]$.
\comment{We varied the scaling factor $\alpha$ in \eqref{eq:pois,gau,stat} to be $0.02:0.005:0.035$
so that the averaged counts per pixel $|\A(\x)|^2 + \b$ 
were approximately from 6 to 25.}
 
\textbf{Implemented Algorithms.}
For unregularized algorithms,
we implemented Gaussian WF, Poisson WF and Poisson-Gaussian WF.
For regularized algorithms,
we implemented 
smoothed total variation (TV)
based on the Huber function~\cite[p.~184]{huber:81}
and PnP/RED methods with the DnCNN denoiser \cite{zhang:17:bag}:
PnP-ADMM~\cite{Venkatakrishnan.etal2013},
PnP-PGM~\cite{ Kamilov.etal2017},  
and RED-SD~\cite{romano:17:tle}.
We also implemented the RED-SD algorithm
with ``Noise2Self" zero-shot image denoising network \cite{batson:19:nbd} 
(RED-SD-SELF).
For diffusion-based models,
we implemented DOLPH~\cite{shoushtari:22:dolph}
and our proposed AWFS.
The Appendix~\ref{appendix, alg} 
shows the implementation details of each algorithm.
We used spectral initialization \cite{luo:19:osi}
for the Gaussian PR and Poisson PR methods;
we then used the output results from Poisson PR
to initialize other algorithms.
We ran all algorithms until convergence in 
normalized root mean squared error (NRMSE) 
or reached the maximum number of iterations (\eg, 50).

To evaluate the robustness and limitation of these algorithms,
we first tuned the parameters for each algorithm
at the noise level when $\alpha=0.030$ and $\sigma=1$,
and then held them fixed throughout all experiments
(\tref{tab:gau,pois,pg}, \tref{tab:pg,reg}, \fref{fig:ssim,nrmse,alpha,sigma}
and \fref{fig:ddpm,asm}).
This is because in practice 
the ground truths are unknown
so that it is impossible to tune parameters 
on each testing data.
Though
the numbers reported can fluctuate 
after careful refinement, 
\eg, by performing grid search on tuning parameters,
such techniques would potentially 
impede the algorithm's practical use.

\textbf{Network Training.}
For PnP denoising networks, we trained all denoisers on different noise levels
$\sigma \in \{9, 11, 13, 15\}$ 
and found that $\sigma=15$ worked the best 
on our data. 
We also used the denoiser scaling technique from~\cite{Xu.etal2020a} 
to dynamically adjust the performance of all PnP methods.
To performing score matching,
we applied 20 geometrically spaced noise levels between 0.005 and 0.1 on each of the training images. 
All networks were implemented in PyTorch
and trained on an NVIDIA Quadro RTX 5000 GPU using the ADAM optimizer \cite{kingma:15:aam}
for 1000 epochs with the best one being selected based off the validation error, \ie, the mean squared error (MSE) loss.
%

\subsection{Results}

We compared all implemented algorithms 
both qualitatively,
by visualizing 
the reconstructed images and residual errors,
and quantitatively,
by computing 
the NRMSE
and structural similarity index measure (SSIM).
Due to the global phase ambiguity,
\ie,
all the algorithms 
can recover the signal only to 
within a constant phase shift
due to the loss of global phase information,
we corrected the phase of $\hat{\x}$ by 
\(
\hat{\x}_{\textup{corrected}} 
\defequ 
\mathrm{sign}\paren{\langle \hat{\x}, \xtrue \rangle}
\hat{\x}.
\)


\begin{figure*}[t]
    \centering\includegraphics[width=\linewidth]{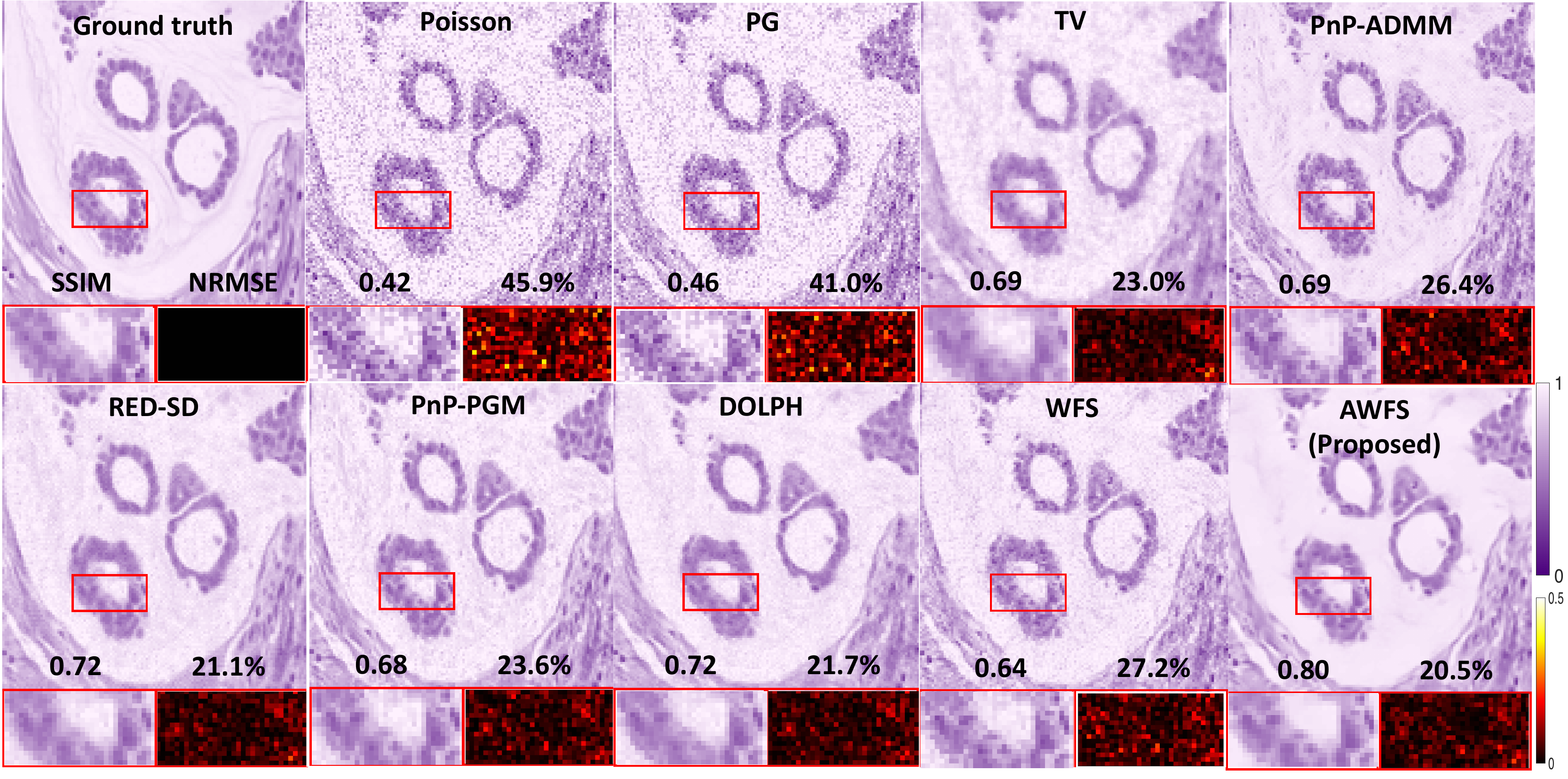}
    \caption{~\emph{Reconstructed images on dataset \cite{aksac:19:bad}. The bottom left/right subfigures correspond to the zoomed in area and the error map for each image. $\alpha$ and $\sigma$ were set to 0.02 and 1, respectively.}}
    \label{fig:visualize,results,breast}
\end{figure*}

\begin{figure*}[ht!]
    \centering\includegraphics[width=\linewidth]{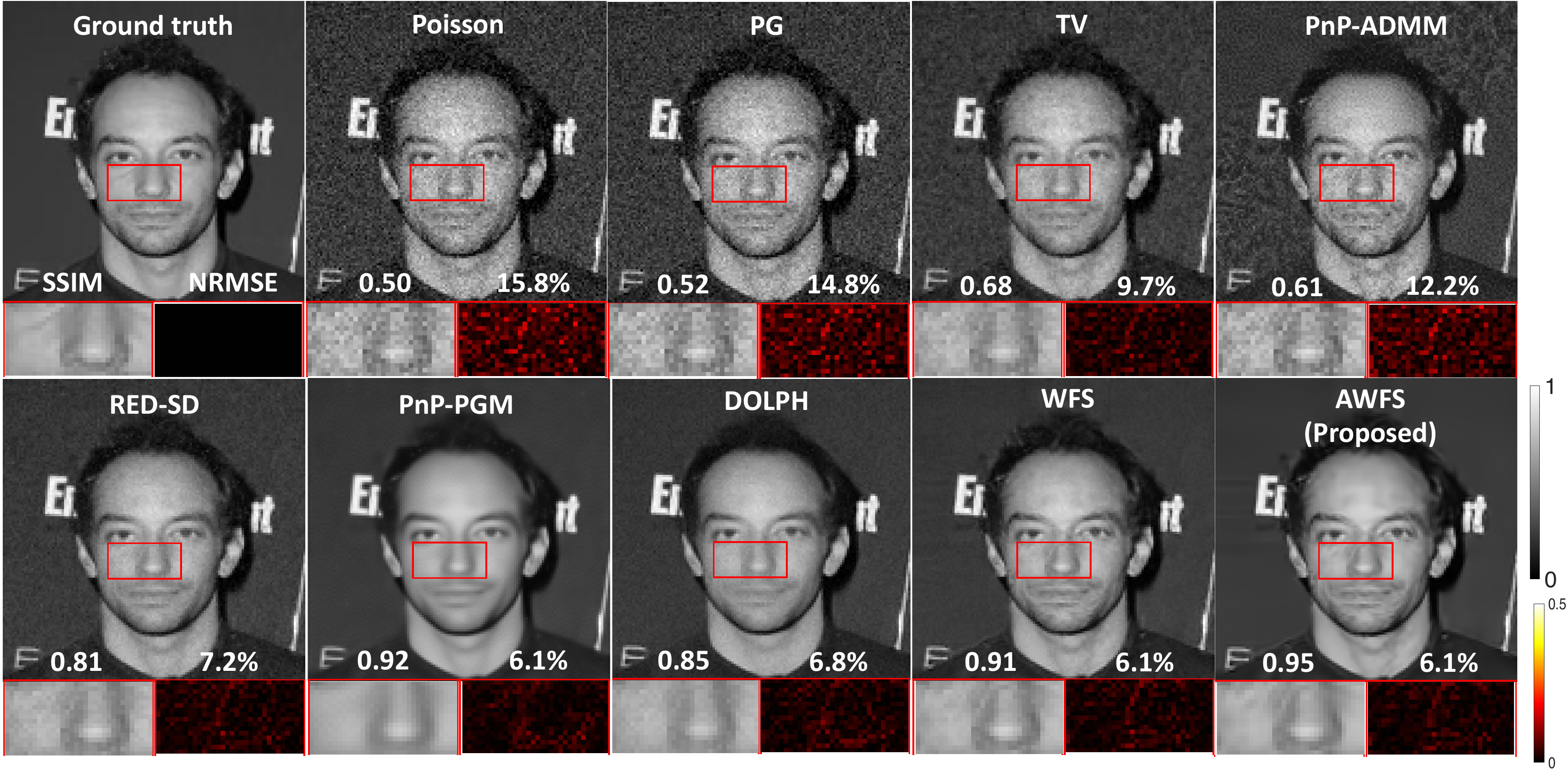}
    \caption{~\emph{Reconstructed images on celebA dataset \cite{liu::15:dlf}. The bottom left/right subfigures correspond to the zoomed in area and the error map for each image. $\alpha$ and $\sigma$ were set to 0.035 and 1, respectively.}}
    \label{fig:visualize,results,celebA}
\end{figure*}

\begin{figure*}[ht!]
    \centering\includegraphics[width=\linewidth]{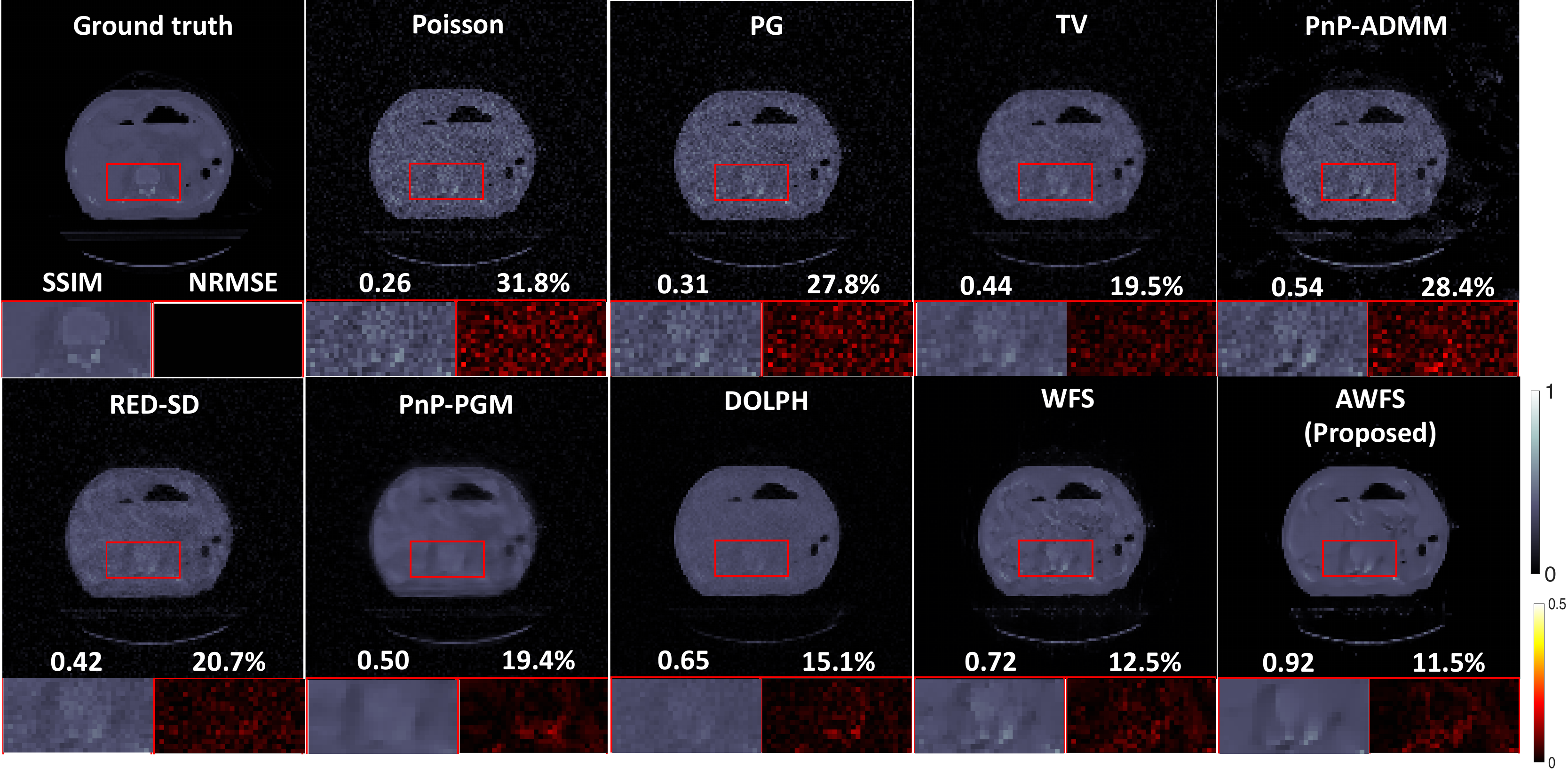}
    \caption{~\emph{Reconstructed images on CT-density dataset. The bottom left/right subfigures correspond to the zoomed in area and the error map for each image. $\alpha$ and $\sigma$ were set to 0.035 and 1, respectively.}}
    \label{fig:visualize,results,density}
\end{figure*}

\begin{table*}[t]
\caption{
SSIM and NRMSE for Poisson and Poisson-Gaussian likelihoods.
Results were averaged across 7 different noise levels 
by varying $\alpha \in 0.02:0.005:0.035$ 
in \eqref{eq:pois,gau,stat}.}

\label{tab:gau,pois,pg}
\centering
\resizebox{\linewidth}{!}{
\begin{tabular}{|c|c|c|c|c|c|c|}
\hline 
Likelihood & 
\multicolumn{2}{c|}{Unregularized (SSIM/NRMSE)}  &  \multicolumn{2}{c|}{DOLPH (SSIM/NRMSE)} & 
\multicolumn{2}{c|}{AWFS (SSIM/NRMSE)}
\\
\hline 
\multicolumn{7}{|c|}{DataSet: Histopathology \cite{aksac:19:bad}}
\\
\hline 
Poisson &
0.54 $\pm$ 0.18
& 31.7 $\pm$ 10.2
& 0.72 $\pm$ 0.13
& 19.5 $\pm$ 6.1
& 0.83 $\pm$ 0.06
& 16.2 $\pm$ 3.7
\\
\hline 
Poisson-Gaussian 
& 0.57 $\pm$ 0.18
& 28.9 $\pm$ 9.0
& 0.80 $\pm$ 0.06
& 16.0 $\pm$ 2.9
& \textbf{0.85 $\pm$ 0.05}
& \textbf{15.4 $\pm$ 3.7}
\\
\hline 
\multicolumn{7}{|c|}{DataSet: CelebA \cite{liu::15:dlf}}
\\
\hline 
Poisson &
0.39 $\pm$ 0.10
& 24.5 $\pm$ 11.4
& 0.61 $\pm$ 0.12
& 15.6 $\pm$ 10.6
& 0.72 $\pm$ 0.16
& 15.2 $\pm$ 11.8
\\
\hline 
Poisson-Gaussian 
& 0.42 $\pm$ 0.10
& 21.8 $\pm$ 9.1
& 0.71 $\pm$ 0.11
& \textbf{13.7 $\pm$ 11.1}
& \textbf{0.74 $\pm$ 0.15}
& 14.8 $\pm$ 11.9
\\
\hline 
\multicolumn{7}{|c|}{DataSet: CT-Density}
\\
\hline 
Poisson &
0.19 $\pm$ 0.06
& 48.9 $\pm$ 13.1
& 0.38 $\pm$ 0.11
& 25.6 $\pm$ 7.5
& 0.84 $\pm$ 0.08
& 17.8 $\pm$ 4.3
\\
\hline 
Poisson-Gaussian 
& 0.24 $\pm$ 0.06
& 40.8 $\pm$ 9.5
& 0.55 $\pm$ 0.08
& 20.0 $\pm$ 3.3
& \textbf{0.88 $\pm$ 0.05}
& \textbf{16.4 $\pm$ 3.7}
\\
\hline 
\end{tabular}
}
\end{table*}

\begin{table*}[t]
\caption{
SSIM and NRMSE using Poisson Gaussian likelihood with different regularization/image prior approaches. 
Results were averaged across 7 different noise levels 
by varying $\alpha \in 0.02:0.005:0.035$ 
in \eqref{eq:pois,gau,stat}.}

\label{tab:pg,reg}
\centering
\resizebox{\linewidth}{!}{
\begin{tabular}{|c|c|c|c|c|c|c|}
\hline 
Dataset &
\multicolumn{2}{c|}{Histopathology \cite{aksac:19:bad}}
&
\multicolumn{2}{c|}{CelebA \cite{liu::15:dlf}}
&
\multicolumn{2}{c|}{CT-Density}
\\
\hline 
Methods & SSIM & NRMSE (\%)
 & SSIM & NRMSE (\%)
 & SSIM & NRMSE (\%)
\\
\hline 
Unregularized
& 0.57 $\pm$ 0.18 
& 28.9 $\pm$ 9.0
& 0.42 $\pm$ 0.10
& 21.8 $\pm$ 9.1
& 0.24 $\pm$ 0.06
& 40.8 $\pm$ 9.5
\\
\hline 
RED-SD-SELF \cite{batson:19:nbd}
& 0.66 $\pm$ 0.13 
& 21.9 $\pm$ 4.5
& 0.60 $\pm$ 0.09
& 15.9 $\pm$ 10.6
& 0.34 $\pm$ 0.04
& 28.1 $\pm$ 4.1
\\
\hline 
PnP-ADMM \cite{Venkatakrishnan.etal2013}
& 0.71 $\pm$ 0.11 
& 20.7 $\pm$ 4.2
& 0.56 $\pm$ 0.08
& 16.7 $\pm$ 8.1
& 0.55 $\pm$ 0.03
& 31.2 $\pm$ 2.7
\\
\hline 
TV regularizer
& 0.72 $\pm$ 0.11 
& 18.2 $\pm$ 3.9
& 0.64 $\pm$ 0.07
& 14.4 $\pm$ 8.6
& 0.41 $\pm$ 0.03
& 23.7 $\pm$ 2.8
\\
\hline 
RED-SD \cite{romano:17:tle}
& 0.76 $\pm$ 0.09
& 16.8 $\pm$ 3.6
& 0.69 $\pm$ 0.11
& 13.9 $\pm$ 10.9
& 0.38 $\pm$ 0.04
& 25.9 $\pm$ 4.0
\\
\hline 
PnP-PGM \cite{Kamilov.etal2017}
& 0.78 $\pm$ 0.11 
& 16.5 $\pm$ 4.5
& \textbf{0.74 $\pm$ 0.14}
& \textbf{13.5 $\pm$ 11.3}
& 0.42 $\pm$ 0.07
& 24.6 $\pm$ 4.4
\\
\hline 
DOLPH \cite{shoushtari:22:dolph}
& 0.80 $\pm$ 0.06 
& 16.0 $\pm$ 2.9
& 0.71 $\pm$ 0.11
& 13.7 $\pm$ 11.1
& 0.55 $\pm$ 0.08
& 20.0 $\pm$ 3.3
\\ 
\hline 
WFS 
& 0.76 $\pm$ 0.12
& 18.2 $\pm$ 5.5
& 0.63 $\pm$ 0.16
& 16.9 $\pm$ 11.8
& 0.53 $\pm$ 0.17
& 21.3 $\pm$ 7.6
\\ 
\hline 
AWFS (Proposed)
& \textbf{0.85 $\pm$ 0.05} 
& \textbf{15.4 $\pm$ 3.7}
& \textbf{0.74 $\pm$ 0.15}
& 14.8 $\pm$ 11.9
& \textbf{0.88 $\pm$ 0.05}
& \textbf{16.4 $\pm$ 3.7}
\\
\hline 
\end{tabular}
}
\end{table*}

\begin{figure*}[ht!]
    \centering
    \subfloat[SSIM varying $\alpha$.]{\includegraphics[width=0.45\linewidth]{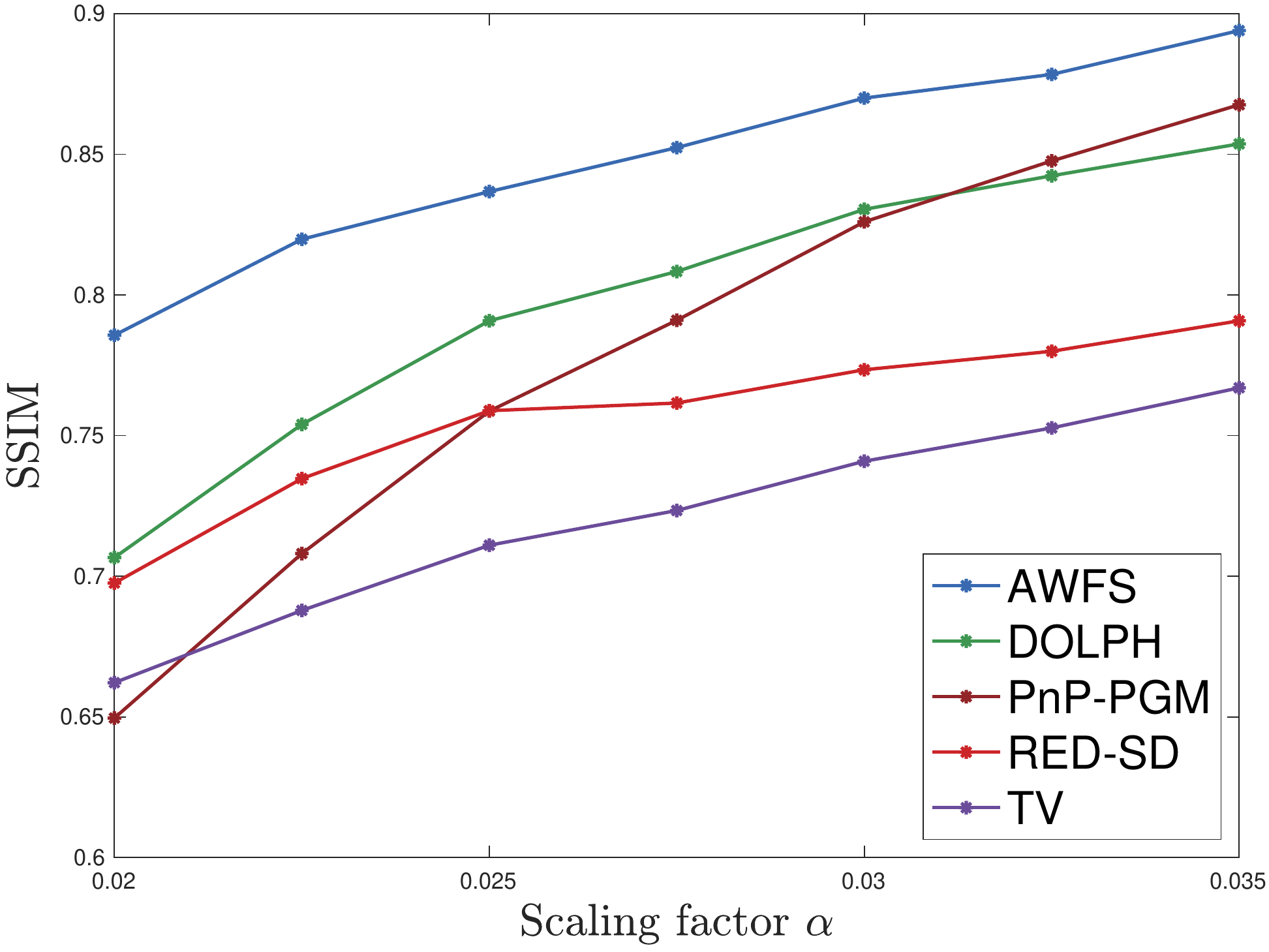}}
    \subfloat[NRMSE varying $\alpha$.]{\includegraphics[width=0.45\linewidth]{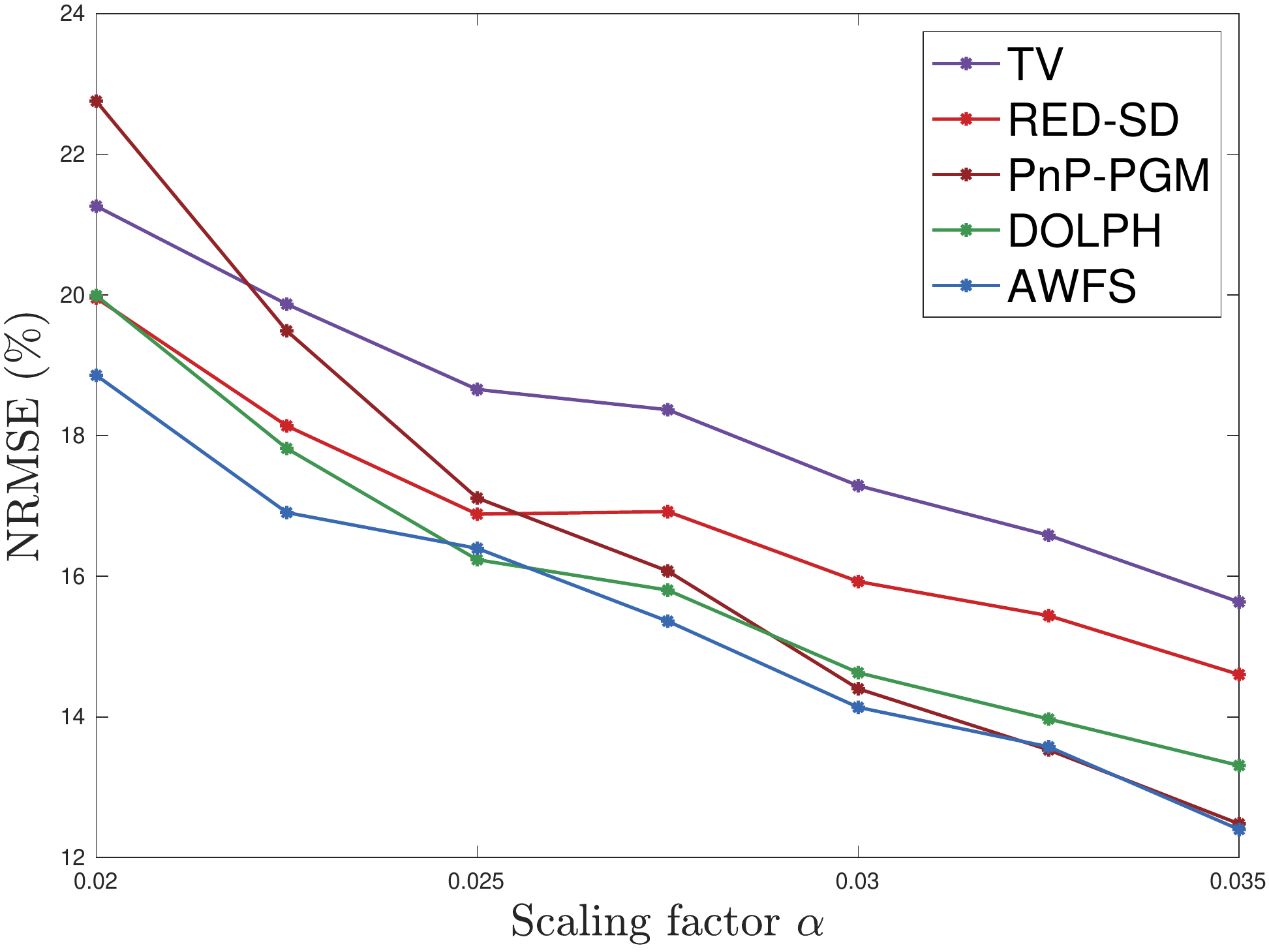}}
    \\
    \subfloat[SSIM varying $\sigma$; $\alpha$ was set as 0.02.]{\includegraphics[width=0.45\linewidth]{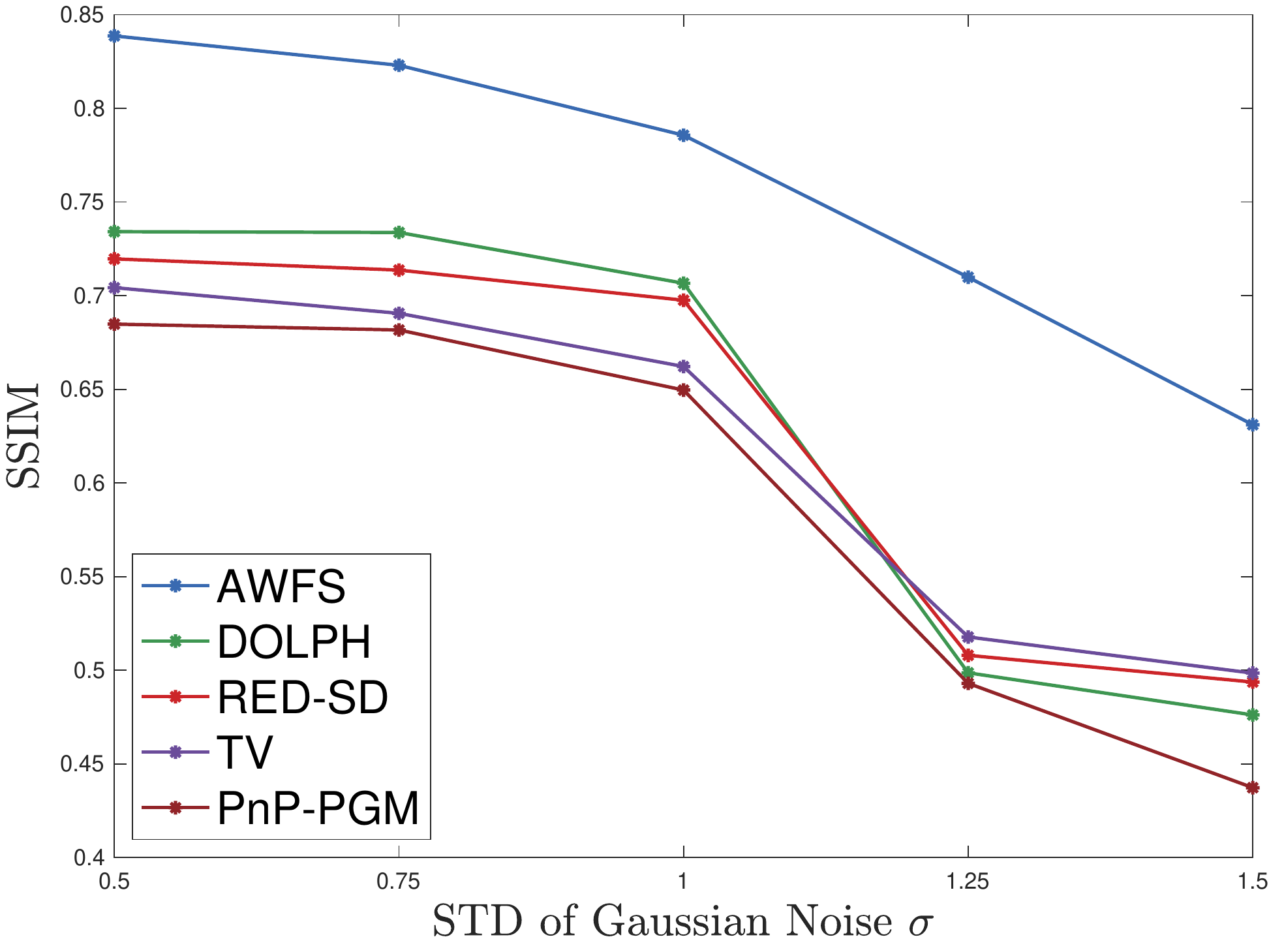}}
    \subfloat[NRMSE varying $\sigma$; $\alpha$ was set as 0.02.]{\includegraphics[width=0.45\linewidth]{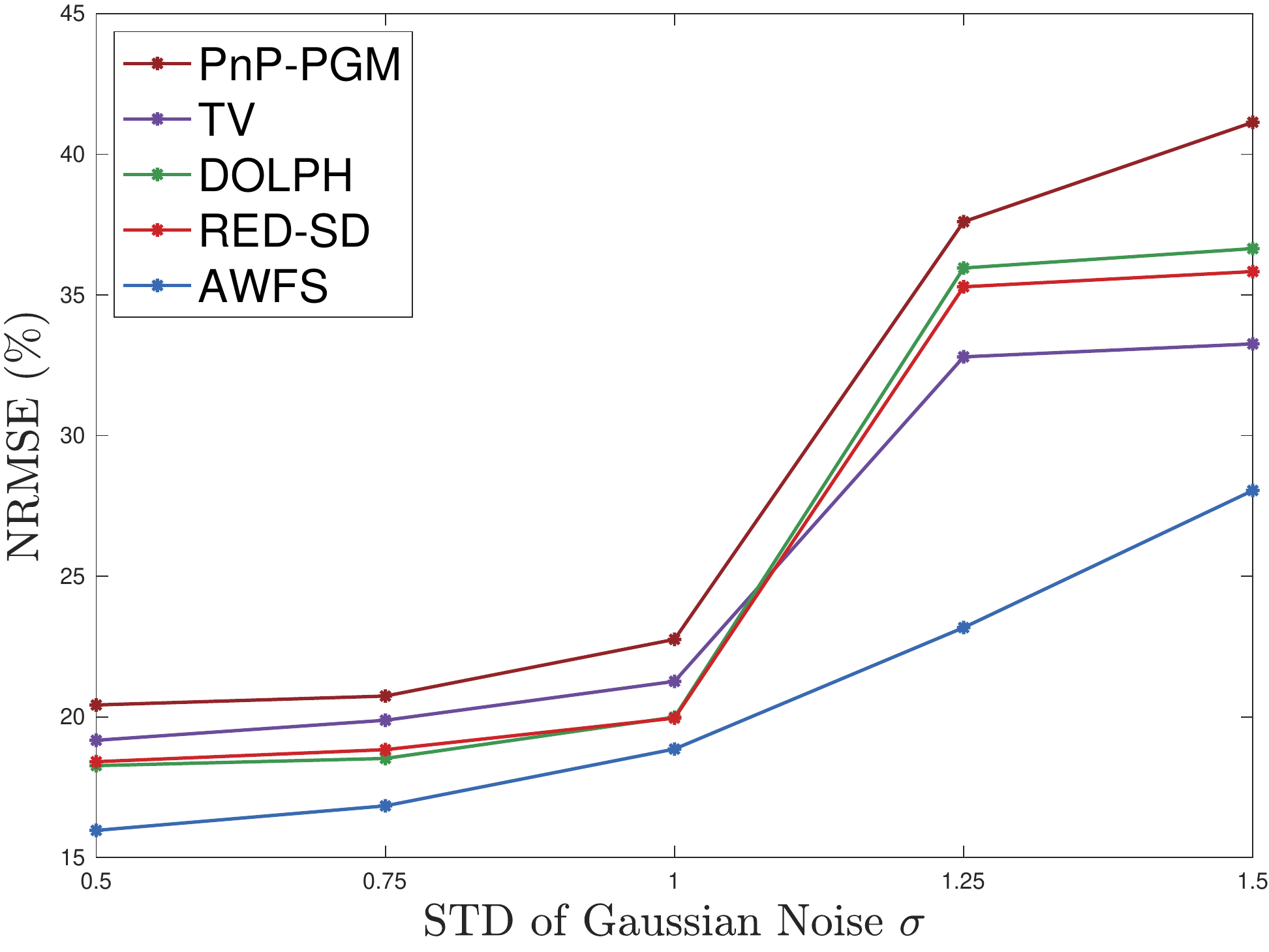}}
    \caption{~\emph{Comparison of SSIM and NRMSE varying scaling factor $\alpha \in [0.02, 0.035]$ 
    and STD of Gaussian noise $\sigma \in [0.25, 1.5]$ defined in \eqref{eq:pois,gau,stat}.}}
    \label{fig:ssim,nrmse,alpha,sigma}
\end{figure*}

\fref{fig:visualize,results,breast},
\fref{fig:visualize,results,celebA}
and \fref{fig:visualize,results,density}
visualize reconstructed images generated by algorithms mentioned in the previous section.
The WF with PG likelihood outperforms 
WF with Poisson likelihood 
with consistently a higher SSIM and lower NRMSE.
Moreover,
we found unregularized Gaussian WF 
failed to reconstruct 
similar to what was reported in \cite{qiu:16:ppr}
(examples are shown in the appendix).
Of the regularized algorithms with PG likelihood, 
our proposed AWFS had less visual noise 
and achieved greater detail recovery compared to other methods, 
as evidenced by the zoomed-in area in these figures.

For quantitative evaluations,
\tref{tab:gau,pois,pg} exemplifies the effect of using our proposed PG likelihood as compared to the simpler Poisson likelihoods. We did not run the Gaussian likelihood with DDPM or AWFS due to the abysmal performance with this likelihood. In all cases, usage of the PG likelihood results in improved image quality in terms of both metrics. 
\tref{tab:pg,reg} consists of experiments using the PG likelihood and shows the efficacy of the proposed AWFS method over other methods.
In particular, our AWFS had superior quantitative performance over all other compared methods on Histopathology and CT-density dataset; in contrast, the PnP-Prox showed the 
lowest NRMSE on celebA dataset. 
This can be due to 
higher randomness in celebrity faces 
because the effectiveness of generative models can vary depending on the dataset being used.
So when provided with a small amount of training data
with high randomness, image denoising models (DnCNN) 
can possibly more effective than generative models.

\comment{
\begin{table}[t]
\caption{
SSIM and NRMSE for different PR methods 
tested on dataset \cite{aksac:19:bad},
averaged across 7 different noise levels 
by varying $\alpha \in 0.02:0.005:0.035$ 
in \eqref{eq:pois,gau,stat}.}

\label{tab:ssim,nrmse}
\centering
\resizebox{0.9\linewidth}{!}{
\begin{tabular}{|c|c|c|c|}
\hline 
Rank & Methods & SSIM $\downarrow$ & NRMSE (\%)
$\uparrow$
\\
\hline 
1 & Poisson-Gaussian with ASM (Proposed)
&0.83 $\pm$ 0.06 
& 16.1 $\pm$ 4.3
\\
\hline 
2 & Poisson with ASM 
& 0.81 $\pm$ 0.07 
& 16.8 $\pm$ 4.3
\\
\hline 
2 & Poisson-Gaussian with DDPM
& 0.79 $\pm$ 0.06 
& 16.4 $\pm$ 3.3
\\
\hline 
4 & Poisson-Gaussian with PnP-PGM
& 0.77 $\pm$ 0.05 
& 17.2 $\pm$ 5.1
\\
\hline 
4 & Poisson-Gaussian with PnP-RED-SD 
& 0.75 $\pm$ 0.09
& 17.0 $\pm$ 3.6
\\
\hline 
6 & Poisson-Gaussian with TV
& 0.71 $\pm$ 0.11 
& 18.7 $\pm$ 4.3
\\
\hline 
7 & Poisson with DDPM
& 0.70 $\pm$ 0.12 
& 20.3 $\pm$ 6.8
\\ 
\hline 
7 & Poisson-Gaussian with PnP-ADMM
& 0.71 $\pm$ 0.10 
& 20.5 $\pm$ 4.1
\\
\hline 
9 & Poisson-Gaussian with RED-SD-SELF
& 0.66 $\pm$ 0.13 
& 22.0 $\pm$ 4.7
\\
\hline 
10 &
Unregularized Poisson-Gaussian 
& 0.55 $\pm$ 0.19 
& 31.8 $\pm$ 11.5
\\
\hline 
11 &
Unregularized Poisson 
& 0.51 $\pm$ 0.19 
& 35.5 $\pm$ 13.2
\\
\hline 
12 
& Unregularized Gaussian 
& 0.28 $\pm$ 0.18 
& 50.6 $\pm$ 12.3
\\
\hline 
\end{tabular}
}
\end{table}
}

\begin{figure*}[t]
    \centering
    \includegraphics[width=\linewidth]{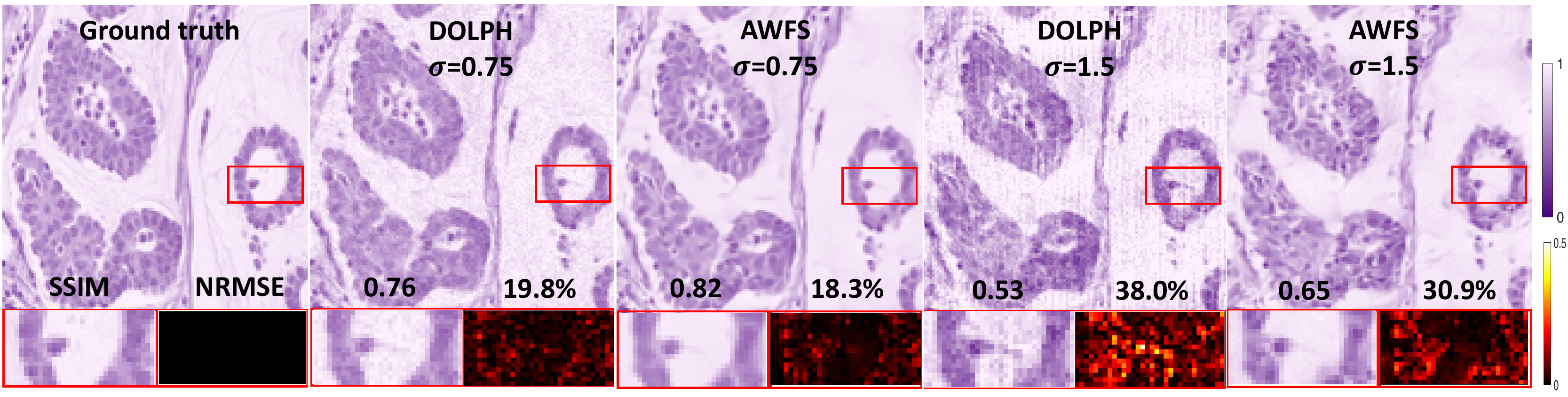}
    \caption{~\emph{Reconstructed images by DOLPH \cite{shoushtari:22:dolph} and 
    our proposed AWFS method under different $\sigma$ values.
    Scaling factor $\alpha$ was set to 0.02 (defined in \eqref{eq:pois,gau,stat}).}}
    \label{fig:ddpm,asm}
\end{figure*}

We also tested
the robustness of the leading algorithms in \tref{tab:pg,reg},
by varying both scaling factor $\alpha$ 
and STD of Gaussian noise $\sigma$.
Results are illustrated in \fref{fig:ssim,nrmse,alpha,sigma} 
and
\fref{fig:ddpm,asm},
where our AWFS algorithm
had the highest SSIM and lowest NRMSE.
In \fref{fig:ddpm,asm},
AWFS demonstrated minimal variations 
in SSIM and NRMSE metrics than DOLPH
as evidenced by the smaller discrepancies in SSIM (0.17 
vs. 0.23) and NRMSE (12.6\% vs. 18.2\%)
when $\sigma$ varies from 0.75 to 1.5.
\fref{fig:uncertainty,plots,score} 
compares the convergence rate of AWF vs. WF 
for the Poisson and PG likelihood, respectively.
Under a variety of noise level,
we found the AWF consistently converges faster
than WF in terms of number of iterations.

\begin{figure*}[ht!]
    \centering
    \subfloat[Histopathology dataset: $\alpha=0.02$.]{\includegraphics[width=0.3\linewidth]{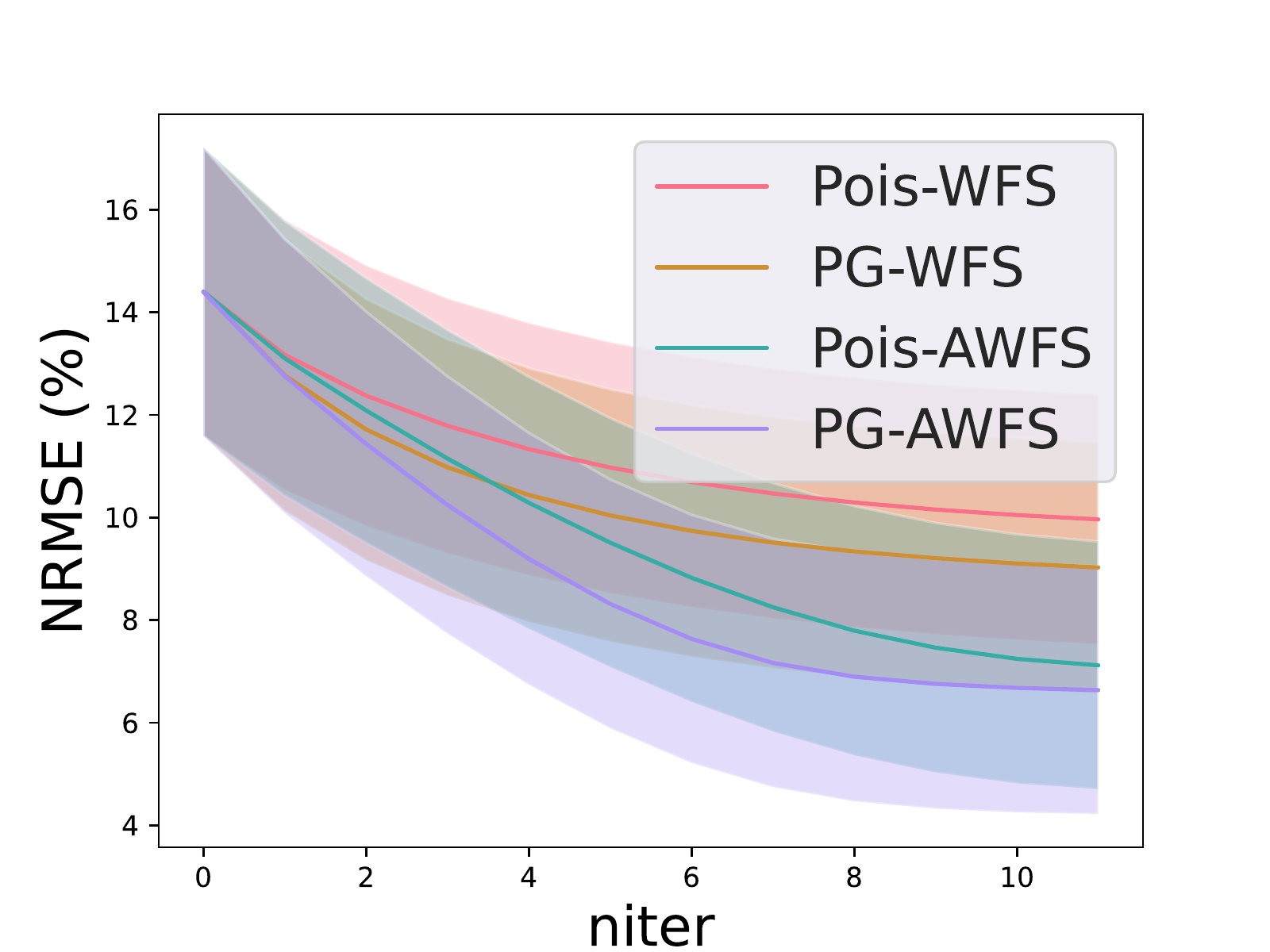}}
    \subfloat[Histopathology dataset: $\alpha=0.0275$.]{
    \includegraphics[width=0.3\linewidth]{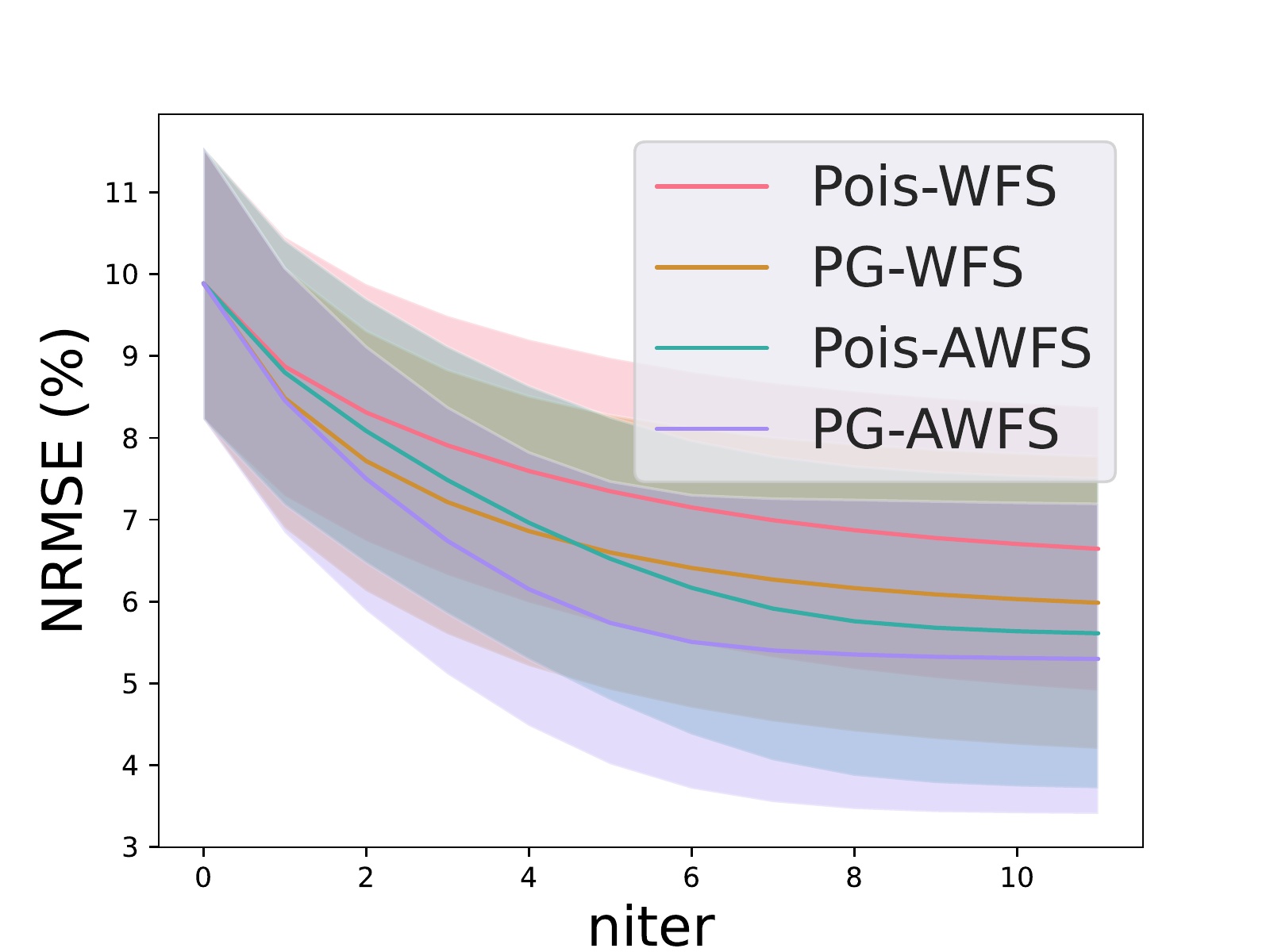}
    }
    \subfloat[Histopathology dataset: $\alpha=0.035$.]{
    \includegraphics[width=0.3\linewidth]{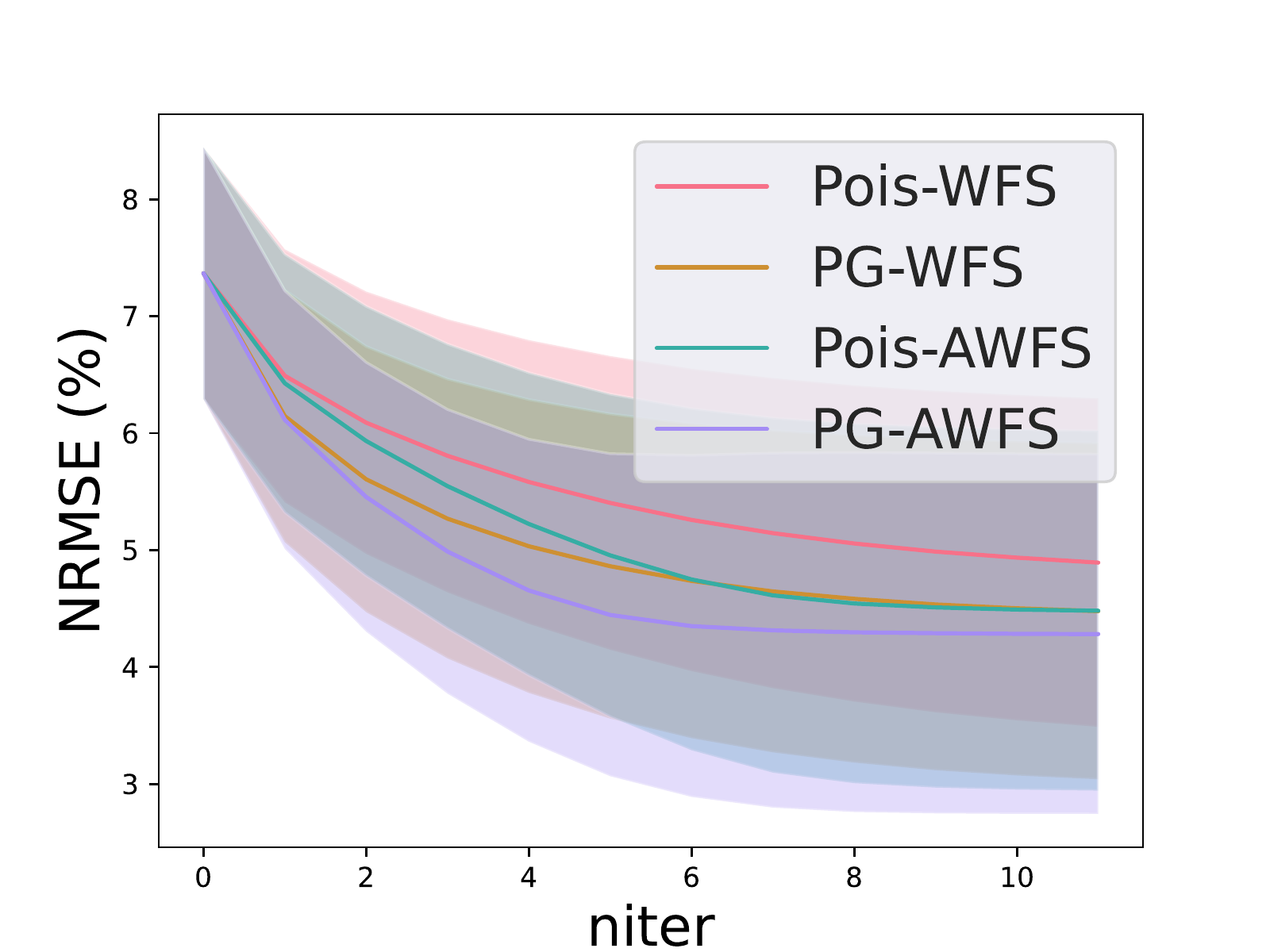}
    }
    \\
    \subfloat[CelebA dataset: $\alpha=0.02$.]{\includegraphics[width=0.3\linewidth]{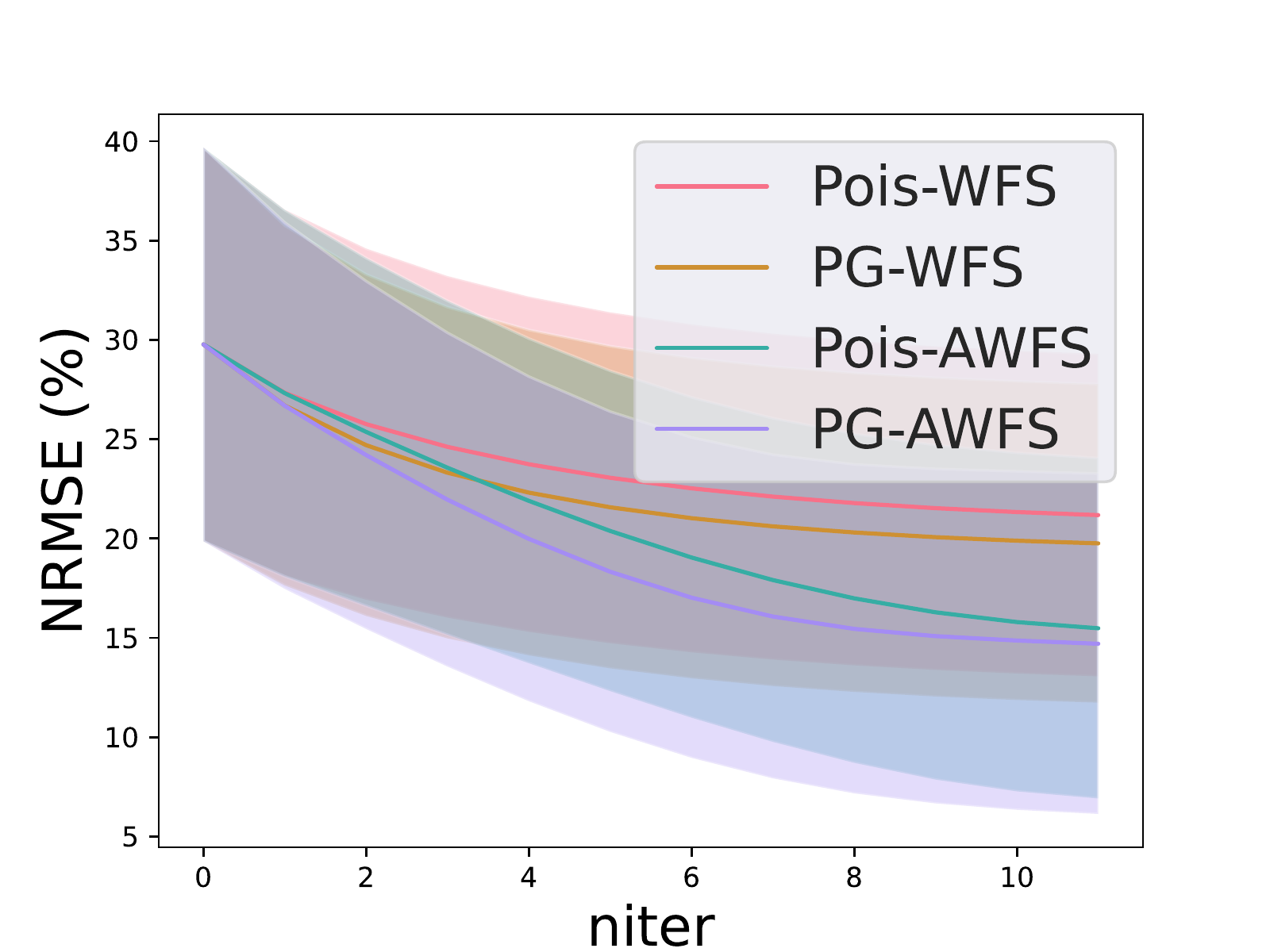}}
    \subfloat[CelebA dataset: $\alpha=0.0275$.]{
    \includegraphics[width=0.3\linewidth]{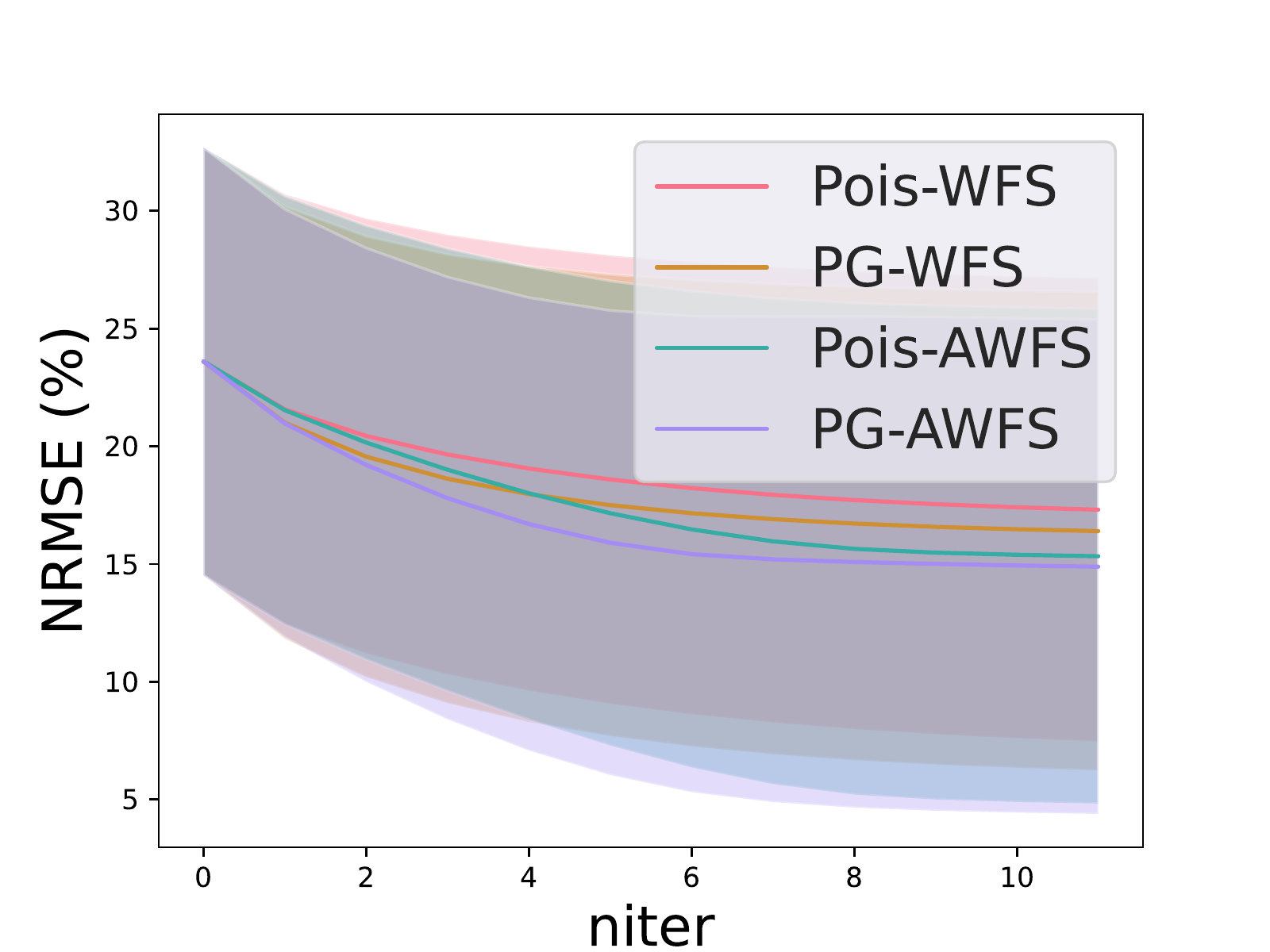}
    }
    \subfloat[CelebA dataset: $\alpha=0.035$.]{
    \includegraphics[width=0.3\linewidth]{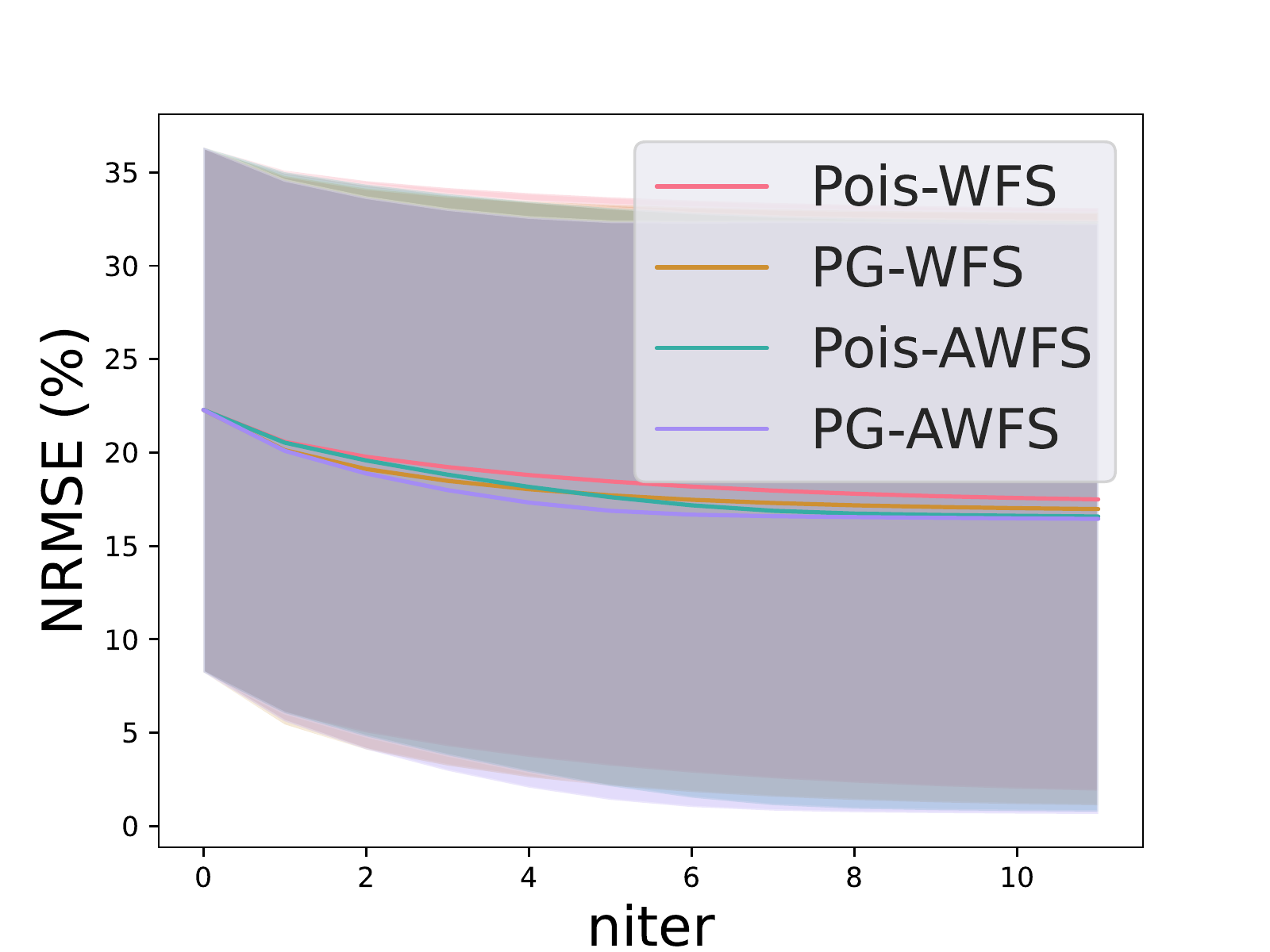}
    }
    \\
    \subfloat[CT-density dataset: $\alpha=0.02$.]{\includegraphics[width=0.3\linewidth]{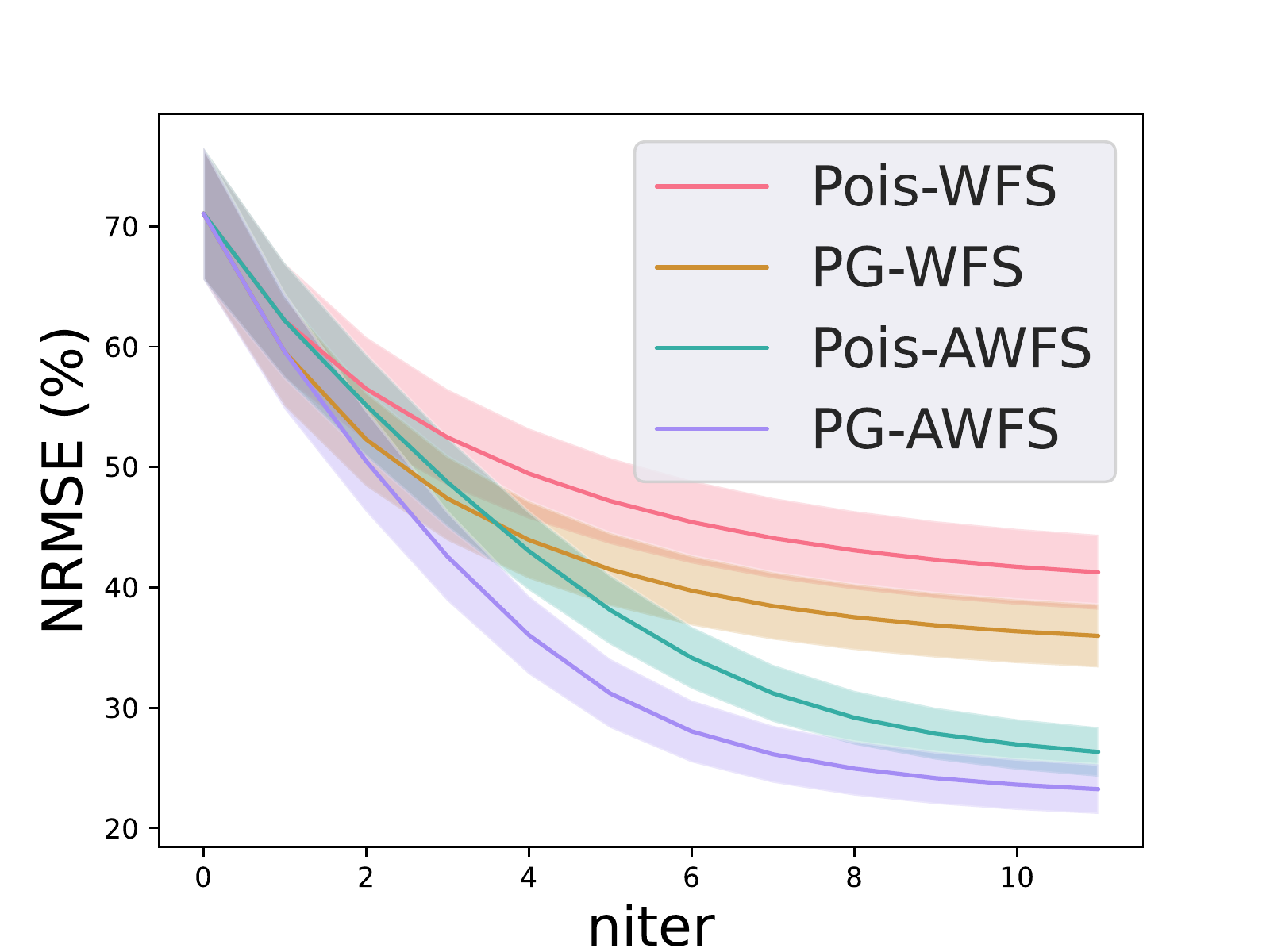}}
    \subfloat[CT-density dataset: $\alpha=0.0275$.]{
    \includegraphics[width=0.3\linewidth]{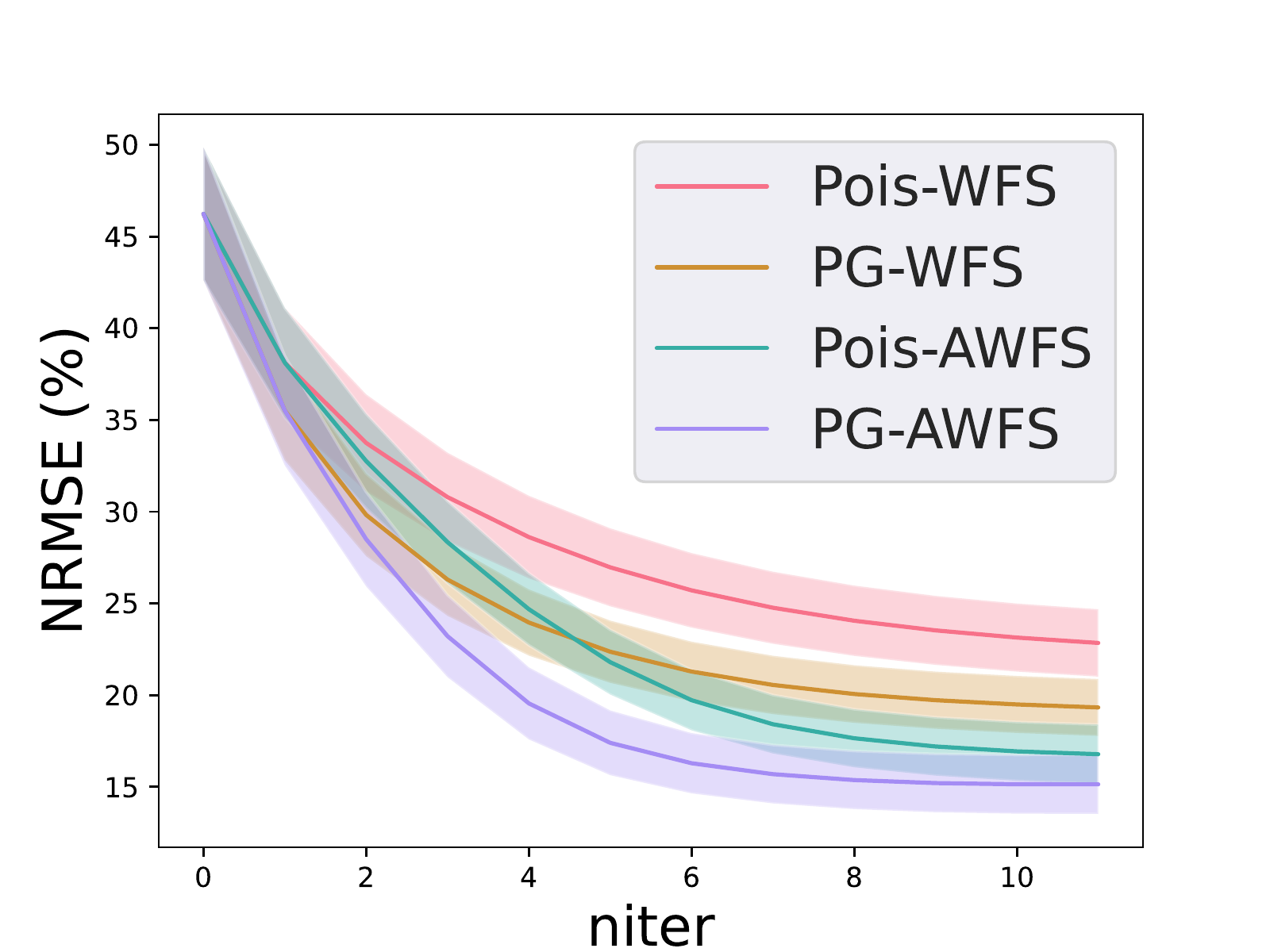}
    }
    \subfloat[CT-density dataset: $\alpha=0.035$.]{
    \includegraphics[width=0.3\linewidth]{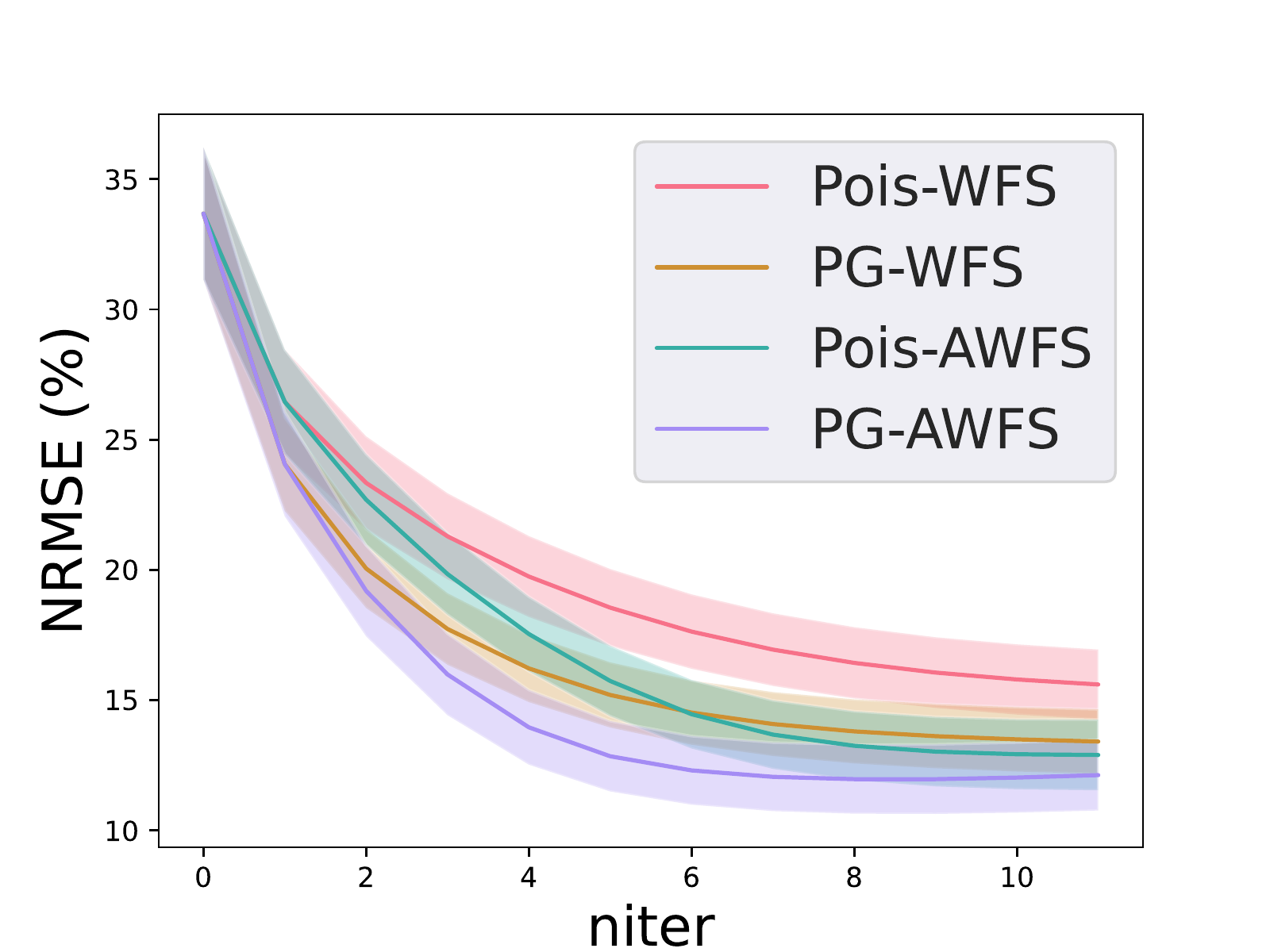}
    }
    \caption{~\emph{Comparing AWF vs. WF with NRMSE vs. number of iterations under different noise levels. The curves and shadows represent the mean and standard deviation, respectively.}}
    \label{fig:uncertainty,plots,score}
\end{figure*}

\section{Discussion}
PR has a long-standing history 
in the field of signal processing and imaging. 
Pioneering works such as  
the error reduction (ER) and
hybrid input-output (HIO) algorithms
by 
Gerchberg Saxton \cite{gerchberg:72:paf}
and Fienup \cite{fienup:82:pra}
have been proposed to address this problem.
These iterative algorithms involve constraints imposed on evaluations between the image domain and frequency domain, 
however, these methods have limitations in terms of the quality of reconstructed images and their convergence remains uncertain \cite{yuan:17:prv}.
Another approach to solving PR problems is through 
compressed sensing and optimization techniques like
Wirtinger flow (WF) \cite{candes:15:prv},
matrix lifting \cite{candes:13:pea, candes:13:prv, shectman:15:prw},
MM \cite{qiu:16:ppr} and ADMM \cite{liang:18:prv}.
In this paper, we focus on the WF algorithm 
due to it is straightforward to
incorporate with the DL regularizer for the image prior.
The likelihood modelling of the noise statistics 
existing in the measurement is also critical. 
Previous studies have primarily focused on modelling either Gaussian or Poisson likelihood only, but in practical scenarios, both types of noise are often encountered.
Therefore, this paper contributes to a more practical perspective of addressing the holographic PR problem by using a PG likelihood and incorporating state-of-the-art deep learning image priors.

In our evaluation of three datasets, 
we consistently observed that the use of PG likelihood yielded superior performance compared to using either Poisson or Gaussian likelihood alone. 
This finding aligns with the natural expectation of accurately modeling the noise likelihood. 
Additionally, we noted that the results obtained from the CT-density dataset were generally in less quality in comparison to those from the other two datasets. 
This can be attributed to lower average counts per pixel 
(plenty of blank pixels near the image boundary).

The use of DL image prior can be considered from two folds: training a denoiser or training to learn the density distribution of images. In our work, 
we applied both approaches and observed that the effectiveness of these methods differs depending on the dataset tested.
Specifically, in the Histopathology dataset \cite{aksac:19:bad}
and the CT-density dataset, 
where the images share similar structures, 
the generative models performs better even trained on
limited amount of data.
In the case of the CelebA dataset \cite{liu::15:dlf}, 
which includes a wide variety of celebrity faces, 
generative models do not exhibit as strong performance as denoiser methods when trained on limited data. 
This is likely due to the fact that generating high-quality images is generally more challenging than removing noise from existing images and may necessitate a larger training dataset.

The effectiveness of accelerated WF compared to the vanilla WF
is due to the nature of non-convexity of the PR problem.
Although recent advances on geometric landscape analysis 
of PR can guarantee that all local minimizers are global
even with random initialization \cite{cai:23:nob},
in practice the measurements are contaminated by noise
so that perhaps much more measurements are required for the
cost function to have a benign geometric landscape. 

Despite the promising results achieved with our proposed AWFS approach, there are several limitations of our work.
First of all, the approximated calculation 
of infinite sum in \eqref{eq:pois,gau,likelihood}, 
though accurate, is computationally expensive.
Therefore, future work should focus on finding ways to accelerate this calculation process while maintaining its accuracy.
Secondly, we did not implement and test the accelerated WF
applied on the diffusion posterior sampling (DPS) method \cite{chung:23:dps}, 
of which the network is fine-tuned from a 
pretrained state-of-the-art diffusion model.
This approach has the potential to advance current methods in PR problem and we will investigate it in the future.

\section{Conclusion}
\label{sec:conclusion}
We proposed a novel algorithm 
based on \textbf{A}ccelerated \textbf{W}irtinger \textbf{F}low 
and \textbf{S}core-based image prior
(AWFS)
for Poisson-Gaussian holographic phase retrieval.
With evaluation on simulated experiments,
we demonstrated that our proposed AWFS
algorithm had the best performance
both qualitatively and quantitatively 
and was more robust to various noise levels,
compared to other state-of-the-art methods.
Furthermore, we proved that 
our proposed algorithm has 
a critical-point convergence guarantee.
Therefore, our approach has much promise 
for translation in real-world applications
encountering phase retrieval problems.


\section*{References}
\printbibliography[heading=none]

@BOOK{huber:81,
 author = {P. J. Huber},
 title = {Robust statistics},
 publisher = {Wiley},
 address = {New York},
 year = 1981
}

@ARTICLE{Makitalo:13:oio,
  author={Makitalo, Markku and Foi, Alessandro},
  journal={IEEE Trans. Imag. Proc.}, 
  title={Optimal Inversion of the Generalized Anscombe Transformation for Poisson-Gaussian Noise}, 
  year={2013},
  volume={22},
  number={1},
  pages={91-103},
  doi={10.1109/TIP.2012.2202675}}

@article{chang:18:tvb, 
place={PHILADELPHIA}, 
title={Total variation–based phase retrieval for Poisson noise removal}, 
volume={11}, 
DOI={10.1137/16M1103270}, 
number={1}, 
journal={SIAM J. Imag. Sci.}, 
publisher={SIAM PUBLICATIONS}, 
author={Chang, Huibin and Lou, Yifei 
and Duan, Yuping and Marchesini, Stefano}, 
year={2018}, 
pages={24–55} }

@ARTICLE{tian:19:fpr,
 author = {X. Tian},
 title = {Fourier ptychographic reconstruction using mixed {Gaussian-Poisson} likelihood with total variation regularisation},
 journal = {{Electronics Letters}},
 volume = 55,
 number = 19,
 pages = {{1041--3}},
 month = sep,
 doi = {10.1049/el.2019.1141},
 year = 2019
}

@ARTICLE{bian:16:fpr,
 author = {L. Bian and J. Suo and J. Chung and X. Ou and C. Yang and F. Chen and Q. Dai},
 title = {Fourier ptychographic reconstruction using {Poisson} maximum likelihood and truncated {Wirtinger} gradient},
 journal = {{Nature Sci. Rep.}},
 volume = 6,
 number = 1,
 doi = {10.1038/srep27384},
 year = 2016
}

@MISC{jaganathan:15:pra,
 author = {K. Jaganathan and Y. C. Eldar and B. Hassibi},
 title = {Phase retrieval: an overview of recent developments},
 note = {\url{https://arxiv.org/abs/1510.07713}},
 year = 2015
}

@ARTICLE{millane:90:pri,
 author = {R. P. Millane},
 title = {Phase retrieval in crystallography and optics},
 journal = {{J. Opt. Soc. Am. A}},
 volume = 7,
 number = 3,
 pages = {{394--411}},
 month = mar,
 doi = {10.1364/JOSAA.7.000394},
 year = 1990
}

@article{dainty:87:pra,
author = {Dainty, J. and Fienup, James},
year = {1987},
month = {01},
pages = {231-275},
title = {Phase retrieval and image reconstruction for astronomy},
volume = {13},
journal = {Imag. Recov. Theory Appl.}
}

@ARTICLE{qiu:16:ppr,
 author = {T. Qiu and P. Babu and D. P. Palomar},
 title = {{PRIME:} phase retrieval via majorization-minimization},
 journal = {{IEEE Trans. Sig. Proc.}},
 volume = 64,
 number = 19,
 pages = {{5174--86}},
 month = oct,
 doi = {10.1109/TSP.2016.2585084},
 year = 2016
}

@article{candes:15:prv,
author = {Candes, Emmanuel and Li, Xiaodong and Soltanolkotabi, Mahdi},
year = {2015},
month = {04},
pages = {1985-2007},
title = {Phase Retrieval via Wirtinger Flow: Theory and Algorithms},
volume = {61},
number = {4},
journal = {IEEE Trans. Info. Theory},
doi = {10.1109/TIT.2015.2399924}
}

@article{gerchberg:72:paf,
  author = {Gerchberg, R. W. and Saxton, W. O.},
  doc-delivery-number = {{M3124}},
  issn = {{0030-4026}},
  journal = {{OPTIK}},
  number = {{2}},
  pages = {{237-246}},
  title = {{Practical Algorithm for Determination of Phase from Image and
  Diffraction Plane Pictures}},
  type = {{Article}},
  unique-id = {{ISI:A1972M312400012}},
  volume = {{35}},
  year = 1972
}

@ARTICLE{thibault:12:mlr,
 author = {P. Thibault and M. Guizar-Sicairos},
 title = {Maximum-likelihood refinement for coherent diffractive imaging},
 journal = {{New J. of Phys.}},
 volume = 14,
 number = 6,
 pages = 063004,
 month = jun,
 doi = {10.1088/1367-2630/14/6/063004},
 year = 2012
}

@ARTICLE{goy:18:lpc,
 author = {A. Goy and K. Arthur and S. Li and G. Barbastathis},
 title = {Low photon count phase retrieval using deep learning},
 journal = {{Phys. Rev. Lett.}},
 volume = 121,
 number = 24,
 pages = 243902,
 month = dec,
 doi = {10.1103/PhysRevLett.121.243902},
 year = 2018
}

@MISC{xu:18:awf,
%  author = {R. Xu and M. Soltanolkotabi and J. P. Haldar and W. Unglaub and J. Zusman and A. F. J. Levi and R. M. Leahy},
%  title = {Accelerated {Wirtinger} flow: {A} fast algorithm for ptychography},
%  url = {http://arxiv.org/abs/1806.05546},
%  year = 2018
% }

@ARTICLE{vazquez:21:qpr,
 author = {I. Vazquez and I. E. Harmon and J. C. R. Luna and M. Das},
 title = {Quantitative phase retrieval with low photon counts using an energy resolving quantum detector},
 journal = {{J. Opt. Soc. Am. A}},
 volume = 38,
 number = 1,
 pages = {{71--9}},
 month = jan,
 doi = {10.1364/JOSAA.396717},
 year = 2021
}

@ARTICLE{candes:13:prv,
 author = {E. J. {Candes} and Y. C. Eldar and T. Strohmer and V. Voroninski},
 title = {Phase retrieval via matrix completion},
 journal = {{SIAM J. Imaging Sci.}},
 volume = 6,
 number = 1,
 pages = {{199--225}},
 doi = {10.1137/110848074},
 year = 2013
}

@ARTICLE{chen:17:srq,
 author = {Y. Chen and E. J. {Candes}},
 title = {Solving random quadratic systems of equations is nearly as easy as solving linear systems},
 journal = {{Comm. Pure Appl. Math.}},
 volume = 70,
 number = 5,
 pages = {{822--83}},
 month = may,
 doi = {10.1002/cpa.21638},
 year = 2017
}

@ARTICLE{zhang:17:fpm,
 author = {Y. Zhang and P. Song and Q. Dai},
 title = {Fourier ptychographic microscopy using a generalized {Anscombe} transform approximation of the mixed {Poisson-Gaussian} likelihood},
 journal = {{Optics Express}},
 volume = 25,
 number = 1,
 pages = {{168--79}},
 month = jan,
 doi = {10.1364/OE.25.000168},
 year = 2017
}

@ARTICLE{zhang:17:ana,
 author = {H. Zhang and Y. Liang and Y. Chi},
 title = {A nonconvex approach for phase retrieval: {Reshaped} {Wirtinger} flow and incremental algorithms},
 journal = {{J. Mach. Learning Res.}},
 volume = 18,
 number = 141,
 pages = {{1--35}},
 url = {http://jmlr.org/papers/v18/16-572.html},
 year = 2017
}

@ARTICLE{snyder:93:irf,
 author = {D. L. Snyder and A. M. Hammoud and R. L. White},
 title = {Image recovery from data acquired with a charge-coupled-device camera},
 journal = {{J. Opt. Soc. Am. A}},
 volume = 10,
 number = 5,
 pages = {{1014--23}},
 month = may,
 doi = {10.1364/JOSAA.10.001014},
 year = 1993
}

@article{shectman:15:prw, 
place={New York}, 
title={Phase Retrieval with Application to Optical Imaging: A contemporary overview}, volume={32}, 
DOI={10.1109/MSP.2014.2352673}, number={3}, journal={IEEE Sig. Proc. Mag.}, publisher={IEEE}, 
author={Shechtman, Yoav and Eldar, Yonina C and Cohen, Oren and Chapman, Henry Nicholas and Jianwei Miao and Segev, Mordechai}, 
year={2015}, 
pages={87–109} }

@article{candes:13:pea, 
title={PhaseLift: Exact and Stable Signal Recovery from Magnitude Measurements via Convex Programming}, 
volume={66}, 
DOI={10.1002/cpa.21432}, 
number={8}, 
journal={Comm. Pure Appl. Math.}, publisher={Hoboken: Wiley Subscription Services, Inc., A Wiley Company}, 
author={Candès, Emmanuel J
and Strohmer, Thomas and Voroninski, Vladislav}, 
pages={1241–1274} ,
year = {2013}
}

@ARTICLE{fatima:22:pam,
  author={Fatima, Ghania and Babu, Prabhu},
  journal={IEEE Sig. Proc. Letters}, 
  title={PGPAL: A monotonic iterative algorithm for Phase-Retrieval under the presence of Poisson-Gaussian noise}, 
  year={2022},
  volume={},
  number={},
  pages={1-1},
  doi={10.1109/LSP.2022.3143469}
}

@article{jiang:16:wfm, 
title={Wirtinger Flow Method With Optimal Stepsize for Phase Retrieval}, 
volume={23}, 
DOI={10.1109/LSP.2016.2611940}, 
number={11}, 
journal={IEEE Sig. Proc. Letters}, 
publisher={IEEE}, 
author={Xue Jiang and  
Rajan, Sreeraman and  
Xingzhao Liu}, 
pages={1627–1631},
year = {2016}
}

@article{li:16:ogd, 
title={On gradient descent algorithm for generalized phase retrieval problem}, 
DOI={10.1109/ICSP.2016.7877848}, 
journal={2016 IEEE 13th International Conference on Signal Processing (ICSP)}, 
publisher={IEEE}, 
author={Li Ji and Zhou Tie}, 
pages={320–325},
year = {2016}
}

@article{gao:17:pru, 
title={Phaseless Recovery Using the Gauss-Newton Method}, volume={65}, 
DOI={10.1109/TSP.2017.2742981}, 
number={22}, 
journal={IEEE Trans. Sig. Proc.}, publisher={IEEE}, 
author={Gao, Bing and Xu, Zhiqiang}, 
pages={5885–5896},
year = {2017}
}

@article{cai:16:oro,
title={Optimal Rates of Convergence for Noisy Sparse Phase Retrieval via Thresholded Wirtinger Flow}, volume={44}, DOI={10.1214/16-AOS1443}, 
number={5}, 
year = {2016},
journal={Annals Stat.}, 
publisher={Hayward: Institute of Mathematical Statistics}, 
author={Cai, T. Tony
and Li, Xiaodong
and Ma, Zongming}, 
pages={2221–2251} 
}

@article{soltanolkotabi:19:ssr, 
title={Structured Signal Recovery From Quadratic Measurements: Breaking Sample Complexity Barriers via Nonconvex Optimization}, 
volume={65}, 
DOI={10.1109/TIT.2019.2891653}, 
number={4}, 
journal={IEEE Trans. Info. Theory}, publisher={IEEE}, 
author={Soltanolkotabi, Mahdi}, 
pages={2374–2400},
year = {2019}
}

@article{netrapalli:15:pru, 
title={Phase Retrieval Using Alternating Minimization}, volume={63}, 
DOI={10.1109/TSP.2015.2448516}, 
number={18}, 
journal={IEEE Trans. Sig. Proc.}, publisher={IEEE}, 
author={Netrapalli, Praneeth
and 
Jain, Prateek
and 
Sanghavi, Sujay}, 
pages={4814–4826},
year = {2015}
}

@article{waldspurger:18:prw, 
title={Phase Retrieval With Random Gaussian Sensing Vectors by Alternating Projections}, 
volume={64}, 
DOI={10.1109/TIT.2018.2800663}, number={5}, 
journal={IEEE Trans. Info. Theory}, publisher={IEEE}, 
author={Waldspurger, Irene}, 
pages={3301–3312},
year = {2018}
}

@article{luo:19:osi, 
title={Optimal Spectral Initialization for Signal Recovery With Applications to Phase Retrieval}, 
volume={67}, 
DOI={10.1109/TSP.2019.2904918}, 
number={9}, 
journal={IEEE Trans. Sig. Proc.}, publisher={IEEE}, 
author={Luo, Wangyu and Alghamdi, Wael and Lu, Yue M}, pages={2347–2356},
year = {2019}
}

@ARTICLE{fatima:21:panp,
  author={Fatima, Ghania and Li, Zongyu and Arora, Aakash and Babu, Prabhu},
  journal={IEEE Trans. Sig. Proc.}, 
  title={PDMM: A novel Primal-Dual Majorization-Minimization algorithm for Poisson Phase-Retrieval problem}, 
  year={2022},
  volume={},
  number={},
  pages={1-1},
  doi={10.1109/TSP.2022.3156014}}

@INPROCEEDINGS{yang:19:pcd,
  author={Yang, Yang and Pesavento, Marius and Eldar, Yonina C. and Ottersten, Björn},
  booktitle={ICASSP 2019 - 2019 IEEE International Conference on Acoustics, Speech and Signal Processing (ICASSP)}, 
  title={Parallel Coordinate Descent Algorithms for Sparse Phase Retrieval}, 
  year={2019},
  volume={},
  number={},
  pages={7670-7674},
  doi={10.1109/ICASSP.2019.8683363}}

@ARTICLE{liang:18:prv,
  author={Liang, Junli and Stoica, Petre and Jing, Yang and Li, Jian},
  journal={IEEE Sig. Proc. Letters}, 
  title={Phase Retrieval via the Alternating Direction Method of Multipliers}, 
  year={2018},
  volume={25},
  number={1},
  pages={5-9},
  doi={10.1109/LSP.2017.2767826}}

@INPROCEEDINGS{li:21:ppr,
  author={Li, Zongyu and Lange, Kenneth and Fessler, Jeffrey A.},
  booktitle={2021 IEEE International Conference on Image Processing (ICIP)}, 
  title={Poisson Phase Retrieval With Wirtinger Flow}, 
  year={2021},
  volume={},
  number={},
  pages={2828-2832},
  doi={10.1109/ICIP42928.2021.9506139}}

@INPROCEEDINGS{sun:16:aga,
  author={Sun, Ju and Qu, Qing and Wright, John},
  booktitle={2016 IEEE International Symposium on Information Theory (ISIT)}, 
  title={A geometric analysis of phase retrieval}, 
  year={2016},
  volume={},
  number={},
  pages={2379-2383},
  doi={10.1109/ISIT.2016.7541725}}

@inproceedings{barmherzig:20:lph,
author = {David A. Barmherzig and Ju Sun},
booktitle = {Imaging and Applied Optics Congress},
journal = {Imaging and Applied Optics Congress},
publisher = {Optica Publishing Group},
title = {Low-Photon Holographie Phase Retrieval},
year = {2020},
doi = {10.1364/COSI.2020.JTu4A.6}
}

@article{latychevskaia:18:ipr,
author = {Tatiana Latychevskaia},
journal = {Appl. Opt.},
number = {25},
pages = {7187--7197},
publisher = {Optica Publishing Group},
title = {Iterative phase retrieval in coherent diffractive imaging: practical issues},
volume = {57},
year = {2018},
doi = {10.1364/AO.57.007187},
}

@INPROCEEDINGS{barmherzig:19:drd,
  author={Barmherzig, David A. and Sun, Ju and Candes, Emmanuel J. and Lane, T.J. and Li, Po-Nan},
  booktitle={2019 13th International conference on Sampling Theory and Applications (SampTA)}, 
  title={Dual-Reference Design for Holographic Phase Retrieval}, 
  year={2019},
  volume={},
  number={},
  pages={1-4},
  doi={10.1109/SampTA45681.2019.9030848}}

@article{barmherzig:19:hpr,
title={Holographic phase retrieval and reference design}, volume={35}, 
DOI={10.1088/1361-6420/ab23d1}, 
number={9}, 
journal={Inverse problems}, 
publisher={Ithaca: IOP Publishing}, 
author={Barmherzig, David A. and Sun, Ju 
and Li, Po-Nan 
and Lane, T.J. 
and Candès, Emmanuel J.}, 
pages={94001-} }

@ARTICLE{li:22:ppr,
  author={Li, Zongyu and Lange, Kenneth and Fessler, Jeffrey A.},
  journal={IEEE Transactions on Computational Imaging}, 
  title={Poisson Phase Retrieval in Very Low-Count Regimes}, 
  year={2022},
  volume={8},
  number={},
  pages={838-850},
  doi={10.1109/TCI.2022.3209936}}

@misc{shoushtari:22:dolph,
      title={DOLPH: Diffusion Models for Phase Retrieval}, 
      author={Shirin Shoushtari and Jiaming Liu and Ulugbek S. Kamilov},
      year={2022},
      eprint={2211.00529},
      archivePrefix={arXiv},
      primaryClass={eess.IV},
}

@INPROCEEDINGS{wu:19:orb,
  author={Wu, Zihui and Sun, Yu and Liu, Jiaming and Kamilov, Ulugbek},
  booktitle={2019 IEEE/CVF International Conference on Computer Vision Workshop (ICCVW)}, 
  title={Online Regularization by Denoising with Applications to Phase Retrieval}, 
  year={2019},
  volume={},
  number={},
  pages={3887-3895},
  doi={10.1109/ICCVW.2019.00482}}

@misc{zhuang:22:ppr,
      title={Practical Phase Retrieval Using Double Deep Image Priors}, 
      author={Zhong Zhuang and David Yang and Felix Hofmann and David Barmherzig and Ju Sun},
      year={2022},
      eprint={2211.00799},
      archivePrefix={arXiv},
      primaryClass={cs.CV}
}

@ARTICLE{zhang:17:bag,
  author={Zhang, Kai and Zuo, Wangmeng and Chen, Yunjin and Meng, Deyu and Zhang, Lei},
  journal={IEEE Transactions on Image Processing}, 
  title={Beyond a Gaussian Denoiser: Residual Learning of Deep CNN for Image Denoising}, 
  year={2017},
  volume={26},
  number={7},
  pages={3142-3155},
  doi={10.1109/TIP.2017.2662206}}

@inproceedings{lehtinen:18:nli,
	author        = {Jaakko Lehtinen and Jacob Munkberg and Jon Hasselgren and Samuli Laine and Tero Karras and Miika Aittala and Timo Aila},
	booktitle     = {Proceedings of the 35th International Conference on Machine Learning},
	url            = {http://proceedings.mlr.press/v80/lehtinen18a.html},
	pages         = {2971--2980},
	publisher     = {PMLR},
	title         = {Noise2Noise: Learning Image Restoration without Clean Data},
	year          = {2018},
}

@InProceedings{batson:19:nbd,
  title = 	 {{N}oise2{S}elf: Blind Denoising by Self-Supervision},
  author =       {Batson, Joshua and Royer, Loic},
  booktitle = 	 {Proceedings of the 36th International Conference on Machine Learning},
  pages = 	 {524--533},
  year = 	 {2019},
  editor = 	 {Chaudhuri, Kamalika and Salakhutdinov, Ruslan},
  volume = 	 {97},
  series = 	 {Proceedings of Machine Learning Research},
  month = 	 {6},
  publisher =    {PMLR},
  url = 	 {https://proceedings.mlr.press/v97/batson19a.html},
}

@INPROCEEDINGS{wang:22:bss,
  author={Wang, Zejin and Liu, Jiazheng and Li, Guoqing and Han, Hua},
  booktitle={2022 IEEE/CVF Conference on Computer Vision and Pattern Recognition (CVPR)}, 
  title={Blind2Unblind: Self-Supervised Image Denoising with Visible Blind Spots}, 
  year={2022},
  volume={},
  number={},
  pages={2017-2026},
  doi={10.1109/CVPR52688.2022.00207}}

@ARTICLE{chan:17:pap,
  author={Chan, Stanley H. and Wang, Xiran and Elgendy, Omar A.},
  journal={IEEE Transactions on Computational Imaging}, 
  title={Plug-and-Play ADMM for Image Restoration: Fixed-Point Convergence and Applications}, 
  year={2017},
  volume={3},
  number={1},
  pages={84-98},
  doi={10.1109/TCI.2016.2629286}}

@ARTICLE{zhang:22:pap,
  author={Zhang, Kai and Li, Yawei and Zuo, Wangmeng and Zhang, Lei and Van Gool, Luc and Timofte, Radu},
  journal={IEEE Transactions on Pattern Analysis and Machine Intelligence}, 
  title={Plug-and-Play Image Restoration With Deep Denoiser Prior}, 
  year={2022},
  volume={44},
  number={10},
  pages={6360-6376},
  doi={10.1109/TPAMI.2021.3088914}}

@article{romano:17:tle,
author = {Romano, Yaniv and Elad, Michael and Milanfar, Peyman},
title = {The Little Engine That Could: Regularization by Denoising (RED)},
journal = {SIAM Journal on Imaging Sciences},
volume = {10},
number = {4},
pages = {1804-1844},
year = {2017},
doi = {10.1137/16M1102884},
}

@inproceedings{gnanasambandam:20:ici,
author = {Gnanasambandam, Abhiram and Chan, Stanley H.},
title = {Image Classification in the Dark Using Quanta Image Sensors},
year = {2020},
url = {https://doi.org/10.1007/978-3-030-58598-3_29},
doi = {10.1007/978-3-030-58598-3_29},
booktitle = {Computer Vision – ECCV 2020: 16th European Conference},
pages = {484–501},
numpages = {18},
}

@InProceedings{lawrence:20:prw,
  title = 	 {Phase Retrieval with Holography and Untrained Priors: Tackling the Challenges of Low-Photon Nanoscale Imaging},
  author =       {Lawrence, Hannah and Barmherzig, David and Li, Henry and Eickenberg, Michael and Gabrie, Marylou},
  booktitle = 	 {Proceedings of the 2nd Mathematical and Scientific Machine Learning Conference},
  pages = 	 {516--567},
  year = 	 {2022},
  volume = 	 {145},
  month = 	 {8},
}

@article{wang:20:piw, 
title={Phase imaging with an untrained neural network}, volume={9}, 
DOI={10.1038/s41377-020-0302-3}, 
number={1}, 
journal={Light, science \& applications},  
author={Wang, F. and Bian, Y. and Wang, H. and others}, 
year = {2020},
pages={77–77}, 
}

@article{zhang:21:pad,
author = {Yuhe Zhang and Mike Andreas Noack and Patrik Vagovic and Kamel Fezzaa and Francisco Garcia-Moreno and Tobias Ritschel and Pablo Villanueva-Perez},
journal = {Opt. Express},
number = {13},
pages = {19593--19604},
publisher = {Optica Publishing Group},
title = {PhaseGAN: a deep-learning phase-retrieval approach for unpaired datasets},
volume = {29},
month = {6},
year = {2021},
url = {https://opg.optica.org/oe/abstract.cfm?URI=oe-29-13-19593},
doi = {10.1364/OE.423222},
}

@article{barmherzig:22:tph,
author = {David A. Barmherzig and Ju Sun},
journal = {Opt. Express},
number = {5},
pages = {6886--6906},
publisher = {Optica Publishing Group},
title = {Towards practical holographic coherent diffraction imaging via maximum likelihood estimation},
volume = {30},
month = {2},
year = {2022},
url = {https://opg.optica.org/oe/abstract.cfm?URI=oe-30-5-6886},
doi = {10.1364/OE.445015},
}

@article{thibault:08:hrs,
author = {Pierre Thibault  and Martin Dierolf  and Andreas Menzel  and Oliver Bunk  and Christian David  and Franz Pfeiffer },
title = {High-Resolution Scanning X-ray Diffraction Microscopy},
journal = {Science},
volume = {321},
number = {5887},
pages = {379-382},
year = {2008},
doi = {10.1126/science.1158573},
URL = {https://www.science.org/doi/abs/10.1126/science.1158573},
eprint = {https://www.science.org/doi/pdf/10.1126/science.1158573}}

@article{marchesini:13:apf,
doi = {10.1088/0266-5611/29/11/115009},
url = {https://dx.doi.org/10.1088/0266-5611/29/11/115009},
year = {2013},
month = {10},
publisher = {IOP Publishing},
volume = {29},
number = {11},
pages = {115009},
author = {Stefano Marchesini and Andre Schirotzek and Chao Yang and Hau-tieng Wu and Filipe Maia},
title = {Augmented projections for ptychographic imaging},
journal = {Inverse Problems},
}

@article{siu:03:uxp,
author = {Karen K-W Siu and Andrei Y. Nikulin and Peter Wells},
title = {Unambiguous x-ray phase retrieval from Fraunhofer diffraction data},
journal = {Journal of Applied Physics},
volume = {93},
number = {5161},
year = {2003},
doi = {https://doi.org/10.1063/1.1565674},
}

@article{chouzenoux:15:aca,
author = {Chouzenoux, Emilie and Jezierska, Anna and Pesquet, Jean-Christophe and Talbot, Hugues},
title = {A Convex Approach for Image Restoration with Exact Poisson--Gaussian Likelihood},
journal = {SIAM Journal on Imaging Sciences},
volume = {8},
number = {4},
pages = {2662-2682},
year = {2015},
doi = {10.1137/15M1014395},
}

@ARTICLE{vincent:11:acb,
  author={Vincent, Pascal},
  journal={Neural Computation}, 
  title={A Connection Between Score Matching and Denoising Autoencoders}, 
  year={2011},
  volume={23},
  number={7},
  pages={1661-1674},
  doi={10.1162/NECO_a_00142}}

@inproceedings{song:19:gmb,
 author = {Song, Yang and Ermon, Stefano},
 booktitle = {Advances in Neural Information Processing Systems},
 pages = {},
 title = {Generative Modeling by Estimating Gradients of the Data Distribution},
 url = {https://proceedings.neurips.cc/paper_files/paper/2019/file/3001ef257407d5a371a96dcd947c7d93-Paper.pdf},
 volume = {32},
 year = {2019}
}

@article{ho:20:ddp,
 author = {Ho, Jonathan and Jain, Ajay and Abbeel, Pieter},
 booktitle = {Advances in Neural Information Processing Systems},
 editor = {H. Larochelle and M. Ranzato and R. Hadsell and M.F. Balcan and H. Lin},
 pages = {6840--6851},
 publisher = {Curran Associates, Inc.},
 title = {Denoising Diffusion Probabilistic Models},
 url = {https://proceedings.neurips.cc/paper_files/paper/2020/file/4c5bcfec8584af0d967f1ab10179ca4b-Paper.pdf},
 volume = {33},
 year = {2020}
}

@inproceedings{dhariwal:21:dmb,
 author = {Dhariwal, Prafulla and Nichol, Alexander},
 booktitle = {Advances in Neural Information Processing Systems},
 pages = {8780--8794},
 title = {Diffusion Models Beat GANs on Image Synthesis},
 url = {https://proceedings.neurips.cc/paper_files/paper/2021/file/49ad23d1ec9fa4bd8d77d02681df5cfa-Paper.pdf},
 volume = {34},
 year = {2021}
}

@inproceedings{song:21:sbg,
  author    = {Yang Song and
               Jascha Sohl{-}Dickstein and
               Diederik P. Kingma and
               Abhishek Kumar and
               Stefano Ermon and
               Ben Poole},
  title     = {Score-Based Generative Modeling through Stochastic Differential Equations},
  booktitle = {9th International Conference on Learning Representations, {ICLR} 2021,
               Virtual Event, Austria, May 3-7, 2021},
  year      = {2021},
  url       = {https://openreview.net/forum?id=PxTIG12RRHS}
}

@inproceedings{jalal:21:rcs,
 author = {Jalal, Ajil and Arvinte, Marius and Daras, Giannis and Price, Eric and Dimakis, Alexandros G and Tamir, Jon},
 booktitle = {Advances in Neural Information Processing Systems},
 pages = {14938--14954},
 publisher = {Curran Associates, Inc.},
 title = {Robust Compressed Sensing MRI with Deep Generative Priors},
 url = {https://proceedings.neurips.cc/paper_files/paper/2021/file/7d6044e95a16761171b130dcb476a43e-Paper.pdf},
 volume = {34},
 year = {2021}
}

@article{chung:22:sbd,
title = {Score-based diffusion models for accelerated MRI},
journal = {Medical Image Analysis},
volume = {80},
pages = {102479},
year = {2022},
issn = {1361-8415},
doi = {https://doi.org/10.1016/j.media.2022.102479},
url = {https://www.sciencedirect.com/science/article/pii/S1361841522001268},
author = {Hyungjin Chung and Jong Chul Ye},
}

@misc{cui:22:sss,
      title={Self-Score: Self-Supervised Learning on Score-Based Models for MRI Reconstruction}, 
      author={Zhuo-Xu Cui and Chentao Cao and Shaonan Liu and Qingyong Zhu and Jing Cheng and Haifeng Wang and Yanjie Zhu and Dong Liang},
      year={2022},
      eprint={2209.00835},
      archivePrefix={arXiv},
      primaryClass={eess.IV}
}

@inproceedings{song:22:sip,
title={Solving Inverse Problems in Medical Imaging with Score-Based Generative Models},
author={Yang Song and Liyue Shen and Lei Xing and Stefano Ermon},
booktitle={International Conference on Learning Representations},
year={2022},
url={https://openreview.net/forum?id=vaRCHVj0uGI}
}

@misc{lee:22:pdo,
      title={Progressive Deblurring of Diffusion Models for Coarse-to-Fine Image Synthesis}, 
      author={Sangyun Lee and Hyungjin Chung and Jaehyeon Kim and Jong Chul Ye},
      year={2022},
      eprint={2207.11192},
      archivePrefix={arXiv},
      primaryClass={cs.CV}
}

@inproceedings{long:09:a3f,
 author = {Y. Long and J. A. Fessler and J. M. Balter},
 title = {A {3D} forward and back-projection method for {X-ray} {CT} using separable footprint},
 booktitle = {{Proc. Intl. Mtg. on Fully 3D Image Recon. in Rad. and Nuc. Med}},
 pages = {{146--9}},
 note = {Winner of poster award.},
 url = {http://web.eecs.umich.edu/~fessler/papers/files/proc/09/web/long-09-a3f.pdf},
 year = 2009
}

@article{aksac:19:bad,
author = {Aksac, A. and Demetrick, D.J. and Ozyer, T. and others},
title = {BreCaHAD: a dataset for breast cancer histopathological annotation and diagnosis},
journal = {MC Res Notes},
volume = {12},
number = {82},
year = {2019},
doi = {https://doi.org/10.1186/s13104-019-4121-7}
}

@article{kamilov2022pnp,
    author={Kamilov, Ulugbek S 
        and Bouman, Charles B 
        and Buzzard, Gregery T 
        and Wohlberg, Brendt},
    title={Plug-and-Play Methods for Integrating 
    Physical and Learned Models in Computational Imaging},
    journal={IEEE Signal Process. Mag.},
    year={2023},
    month={1},
    volume={40},
    number={1},
    pages={85--97}
}

@inproceedings{Venkatakrishnan.etal2013,
  title = {Plug-and-Play Priors for Model Based Reconstruction},
  booktitle = {Proc. {{IEEE}} Global Conf. {{Signal}} Process. and Inf. {{Process}}.},
  author = {Venkatakrishnan, S. V. and Bouman, C. A. and Wohlberg, B.},
  year = {2013},
  month = dec,
  pages = {945--948},
  address = {{Austin, TX, USA}}
}

@article{Kamilov.etal2017,
  title = {A Plug-and-Play Priors Approach for Solving Nonlinear Imaging Inverse Problems},
  author = {Kamilov, U. S. and Mansour, H. and Wohlberg, B.},
  year = {2017},
  month = dec,
  journal = {IEEE Signal. Proc. Let.},
  volume = {24},
  number = {12},
  pages = {1872--1876}
}

@article{Romano.etal2017,
  title = {The Little Engine That Could: {{Regularization}} by Denoising ({{RED}})},
  author = {Romano, Y. and Elad, M. and Milanfar, P.},
  year = {2017},
  journal = {SIAM Journal on Imaging Sciences},
  volume = {10},
  number = {4},
  pages = {1804--1844}
}

@article{Reehorst.Schniter2019,
  title = {Regularization by Denoising: {{Clarifications}} and New Interpretations},
  author = {Reehorst, E. T. and Schniter, P.},
  year = {2019},
  month = mar,
  journal = {IEEE Trans. Comput. Imag.},
  volume = {5},
  number = {1},
  pages = {52--67}
}

@article{Sreehari.etal2016,
  title = {Plug-and-Play Priors for Bright Field Electron Tomography and Sparse Interpolation},
  author = {Sreehari, S. and Venkatakrishnan, S. V. and Wohlberg, B. and Buzzard, G. T. and Drummy, L. F. and Simmons, J. P. and Bouman, C. A.},
  year = {2016},
  month = dec,
  journal = {IEEE Transactions on Computational Imaging},
  volume = {2},
  number = {4},
  pages = {408--423}
}

@article{Ahmad.etal2020,
  ids = {Ahmad.etal2019},
  title = {Plug-and-Play Methods for Magnetic Resonance Imaging: {{Using}} Denoisers for Image Recovery},
  author = {Ahmad, R. and Bouman, C. A. and Buzzard, G. T. and Chan, S. and Liu, S. and Reehorst, E. T. and Schniter, P.},
  year = {2020},
  journal = {IEEE Signal Processing Magazine},
  volume = {37},
  number = {1},
  pages = {105--116}
}

@inproceedings{Zhang.etal2017a,
  title = {Learning Deep {{CNN}} Denoiser Prior for Image Restoration},
  booktitle = {Proc. {{IEEE}} Conf. {{Computer}} Vision and Pattern Recognition},
  author = {Zhang, K. and Zuo, W. and Gu, S. and Zhang, L.},
  year = {2017},
  month = {7},
  pages = {3929--3938}
}

@inproceedings{Zhang.etal2019,
  title = {Deep Plug-and-Play Super-Resolution for Arbitrary Blur Kernels},
  booktitle = {Proceedings of the {{IEEE}} Conference on Computer Vision and Pattern Recognition},
  author = {Zhang, K. and Zuo, W. and Zhang, L.},
  year = {2019},
  pages = {1671--1681},
  address = {{Long Beach, CA, USA}}
}

@inproceedings{Xu.etal2020a,
  title = {Boosting the Performance of Plug-and-Play Priors via Denoiser Scaling},
  booktitle = {54th {{Asilomar Conf}}. on {{Signals}}, {{Systems}}, and {{Computers}}},
  author = {Xu, Xiaojian and Liu, Jiaming and Sun, Yu and Wohlberg, Brendt and Kamilov, Ulugbek S.},
  year = {2020},
  pages = {1305--1312},
  doi = {10.1109/IEEECONF51394.2020.9443410}
}

@article{nesterov:05:smo,
author = {Nesterov, Y.},
title = {Smooth Minimization of Non-Smooth Functions},
year = {2005},
issue_date = {May 2005},
publisher = {Springer-Verlag},
address = {Berlin, Heidelberg},
volume = {103},
number = {1},
issn = {0025-5610},
url = {https://doi.org/10.1007/s10107-004-0552-5},
doi = {10.1007/s10107-004-0552-5},
journal = {Math. Program.},
month = {5},
pages = {127–152},
numpages = {26},
}

@article{kim:16:ofo,
author = {Kim, D. and Fessler, J. A.},
title = {Optimized first-order methods for smooth convex minimization},
journal = {Math Program},
year = {2016},
volume = {159},
number = {1},
page = {81-107},
doi = {10.1007/s10107-015-0949-3},
}

@book{goodfellow:16:dl,
    title={Deep Learning},
    author={Ian Goodfellow and Yoshua Bengio and Aaron Courville},
    publisher={MIT Press},
    note={\url{http://www.deeplearningbook.org}},
    year={2016}
}

@article{ghadimi:16:agm,
author = {Ghadimi, S. and Lan, G},
title = {Accelerated gradient methods for nonconvex nonlinear and stochastic programming},
journal = {Math. Program.},
volume = {156},
page = {59-99},
year = {2016},
doi = {https://doi.org/10.1007/s10107-015-0871-8}
}

@inproceedings{li:15:apg,
 author = {Li, Huan and Lin, Zhouchen},
 booktitle = {Advances in Neural Information Processing Systems},
 title = {Accelerated Proximal Gradient Methods for Nonconvex Programming},
 url = {https://proceedings.neurips.cc/paper_files/paper/2015/file/f7664060cc52bc6f3d620bcedc94a4b6-Paper.pdf},
 volume = {28},
 year = {2015}
}

@article{saliba:12:fth,
author = {M. Saliba and T. Latychevskaia and J. Longchamp and H. Fink},
title = {Fourier Transform Holography: A Lensless Non-Destructive Imaging Technique},
journal = {Microsc. Microanal.},
volume = {18},
number = {S2},
page = {564-565},
year = {2012}
}

@article{ongie:20:dlt,
  title={Deep Learning Techniques for Inverse Problems in Imaging},
  author={Greg Ongie and Ajil Jalal and Christopher A. Metzler and Richard Baraniuk and Alexandros G. Dimakis and Rebecca M. Willett},
  journal={IEEE Journal on Selected Areas in Information Theory},
  year={2020},
  volume={1},
  pages={39-56}
}

@inproceedings{
graikos:22:dma,
title={Diffusion Models as Plug-and-Play Priors},
author={Alexandros Graikos and Nikolay Malkin and Nebojsa Jojic and Dimitris Samaras},
booktitle={Advances in Neural Information Processing Systems},
year={2022},
url={https://openreview.net/forum?id=yhlMZ3iR7Pu}
}

@INPROCEEDINGS{bostan:18:awf,
  author={Bostan, Emrah and Soltanolkotabi, Mahdi and Ren, David and Waller, Laura},
  booktitle={2018 25th IEEE International Conference on Image Processing (ICIP)}, 
  title={Accelerated Wirtinger Flow for Multiplexed Fourier Ptychographic Microscopy}, 
  year={2018},
  volume={},
  number={},
  pages={3823-3827},
  doi={10.1109/ICIP.2018.8451437}}

@INPROCEEDINGS{fabian:20:3pr,
  author={Fabian, Zalan and Haldar, Justin and Leahy, Richard and Soltanolkotabi, Mahdi},
  booktitle={2020 28th European Signal Processing Conference (EUSIPCO)}, 
  title={3D Phase Retrieval at Nano-Scale via Accelerated Wirtinger Flow}, 
  year={2021},
  volume={},
  number={},
  pages={2080-2084},
  doi={10.23919/Eusipco47968.2020.9287703}}

@article{gao:22:psr,
author = {Gao, Y. and Yang, F. and Cao, L.},
title = {Pixel Super-Resolution Phase Retrieval for Lensless On-Chip Microscopy via Accelerated Wirtinger Flow},
journal = {Cells},
year = {2022},
volume = {11},
number = {13},
doi = {10.3390/cells11131999}
}

@article{ye:22:set,
author = {Qiuliang Ye and Li-Wen Wang and Daniel P. K. Lun},
journal = {Opt. Express},
number = {18},
pages = {31937--31958},
publisher = {Optica Publishing Group},
title = {SiSPRNet: end-to-end learning for single-shot phase retrieval},
volume = {30},
year = {2022},
url = {https://opg.optica.org/oe/abstract.cfm?URI=oe-30-18-31937},
doi = {10.1364/OE.464086},
}

@inproceedings{kingma:15:aam,
  author       = {Diederik P. Kingma and
                  Jimmy Ba},
  title        = {Adam: {A} Method for Stochastic Optimization},
  booktitle    = {3rd International Conference on Learning Representations, {ICLR} 2015,
                  San Diego, CA, USA, May 7-9, 2015, Conference Track Proceedings},
  year         = {2015},
  url          = {http://arxiv.org/abs/1412.6980},
  bibsource    = {dblp computer science bibliography, https://dblp.org}
}

@article{guizar-sicairos:07:hwe,
author = {Manuel Guizar-Sicairos and James R. Fienup},
journal = {Opt. Express},
number = {26},
pages = {17592--17612},
publisher = {Optica Publishing Group},
title = {Holography with extended reference by autocorrelation linear differential operation},
volume = {15},
year = {2007},
url = {https://opg.optica.org/oe/abstract.cfm?URI=oe-15-26-17592},
}

@ARTICLE{wang:19:tpc,
  author={Wang, Ligong},
  journal={IEEE Transactions on Information Theory}, 
  title={The Poisson Channel With Varying Dark Current Known to the Transmitter}, 
  year={2019},
  volume={65},
  number={8},
  pages={4966-4978},
  doi={10.1109/TIT.2019.2911474}}

@InProceedings{bora:17:csu,
  title = 	 {Compressed Sensing using Generative Models},
  author =       {Ashish Bora and Ajil Jalal and Eric Price and Alexandros G. Dimakis},
  booktitle = 	 {Proceedings of the 34th International Conference on Machine Learning},
  pages = 	 {537--546},
  year = 	 {2017},
  volume = 	 {70},
  series = 	 {Proceedings of Machine Learning Research},
  publisher =    {PMLR},
}

@article{asim:20:igm,
title = {Invertible generative models for inverse problems: mitigating representation error and dataset bias}, 
url = {https://par.nsf.gov/biblio/10252230}, abstractNote = {}, 
journal = {Proceedings of the 37th International Conference on Machine Learning}, 
volume = {119}, 
year = {2020},
author = {Asim, Muhammad and Daniels, Max and Leong, Oscar and Ahmed, Ali and Hand, Paul}, }

@ARTICLE{wei:22:duw,
  author={Wei, Xinyi and van Gorp, Hans and Gonzalez-Carabarin, Lizeth and Freedman, Daniel and Eldar, Yonina C. and van Sloun, Ruud J. G.},
  journal={IEEE Transactions on Signal Processing}, 
  title={Deep Unfolding With Normalizing Flow Priors for Inverse Problems}, 
  year={2022},
  volume={70},
  number={},
  pages={2962-2971},
  doi={10.1109/TSP.2022.3179807}}

@book{daubechies:92:tlo,
author = {Daubechies, Ingrid},
title = {Ten Lectures on Wavelets},
publisher = {Society for Industrial and Applied Mathematics},
year = {1992},
doi = {10.1137/1.9781611970104},
address = {},
edition   = {},
URL = {https://epubs.siam.org/doi/abs/10.1137/1.9781611970104},
eprint = {https://epubs.siam.org/doi/pdf/10.1137/1.9781611970104}
}

@book{luenberger:97:obv,
author = {Luenberger, David G.},
title = {Optimization by Vector Space Methods},
year = {1997},
isbn = {047155359X},
address = {USA},
edition = {1st},
}

@book{grafakos:04:cam,
author = {Loukas Grafakos},
title = {Classical and Modern Fourier Analysis},
publisher = {Pearson/Prentice Hall},
year = {2004},
}

@inproceedings{chung:22:idm,
 author = {Chung, Hyungjin and Sim, Byeongsu and Ryu, Dohoon and Ye, Jong Chul},
 booktitle = {Advances in Neural Information Processing Systems},
 pages = {25683--25696},
 title = {Improving Diffusion Models for Inverse Problems using Manifold Constraints},
 url = {https://proceedings.neurips.cc/paper_files/paper/2022/file/a48e5877c7bf86a513950ab23b360498-Paper-Conference.pdf},
 volume = {35},
 year = {2022}
}

@inproceedings{
chung:23:dps,
title={Diffusion Posterior Sampling for General Noisy Inverse Problems},
author={Hyungjin Chung and Jeongsol Kim and Michael Thompson Mccann and Marc Louis Klasky and Jong Chul Ye},
booktitle={The Eleventh International Conference on Learning Representations },
year={2023},
url={https://openreview.net/forum?id=OnD9zGAGT0k}
}

@inproceedings{liu::15:dlf,
  title = {Deep Learning Face Attributes in the Wild},
  author = {Liu, Ziwei and Luo, Ping and Wang, Xiaogang and Tang, Xiaoou},
  booktitle = {Proceedings of International Conference on Computer Vision (ICCV)},
  month = {12},
  year = {2015} 
}

@article{li:22:dad, 
    title={DblurDoseNet: A deep residual learning network for voxel radionuclide dosimetry compensating for single‐Photon Emission Computerized Tomography Imaging resolution}, 
    volume={49}, 
    DOI={10.1002/mp.15397}, 
    number={2}, 
    journal={Medical Physics}, 
    author={Li, Zongyu and Fessler, Jeffrey A. and Mikell, Justin K. and Wilderman, Scott J. and Dewaraja, Yuni K.}, 
    year={2022}, 
    pages={1216–1230}
}

@article{fienup:82:pra,
title={Phase Retrieval Algorithms: A Comparison},
author = {J.R. Fienup},
journal = {Appl. Opt.},
volume = {21},
page = {2758-2769},
year = {1982}
}

@article{yuan:17:prv,
author = {Ziyang Yuan and Hongxia Wang},
journal = {Appl. Opt.},
number = {9},
pages = {2418--2427},
publisher = {Optica Publishing Group},
title = {Phase retrieval via reweighted Wirtinger flow},
volume = {56},
month = {3},
year = {2017},
url = {https://opg.optica.org/ao/abstract.cfm?URI=ao-56-9-2418},
doi = {10.1364/AO.56.002418}
}

@article{cai:23:nob,
doi = {10.1088/1361-6420/acdab7},
url = {https://dx.doi.org/10.1088/1361-6420/acdab7},
year = {2023},
month = {6},
publisher = {IOP Publishing},
volume = {39},
number = {7},
pages = {075011},
author = {Jian-Feng Cai and Meng Huang and Dong Li and Yang Wang},
title = {Nearly optimal bounds for the global geometric landscape of phase retrieval},
journal = {Inverse Problems}
}
\clearpage
\newpage
\onecolumn
\appendix 
\setcounter{equation}{0}
\setcounter{figure}{0}
\setcounter{algorithm}{0}
\setcounter{subsection}{0}
\renewcommand{\thefigure}{A.\arabic{figure}}
\renewcommand{\theequation}{A.\arabic{equation}}
\renewcommand{\thealgorithm}{A.\arabic{algorithm}}
\label{sec:append}
This is the appendix for the paper ``Poisson-Gaussian Holographic Phase Retrieval
with Score-based Image Prior''. 

\renewcommand{\thesection}{A}
\comment{
\subsection{Proof of Theorem 1}
\label{appendix, p1}
\comment{
\textbf{Lemma 1.1:}\label{lemma,lips}
The function $\phi(u)$ is 
Lipschitz differentiable 
($\phi$ is defined in \eqref{eq:pois,gau,grad}).
The Lipschitz constant for $\dot\phi(u)$ is:
\begin{equation}\label{eq:lemma,lips}
\max\{|\ddot{\phi}(u)|\} \defequ \mu = \paren{1 - e^{-\frac{1}{\sigma^2}}}
e^{\frac{2 y_{\max} - 1}{\sigma^2}
},
\quad 
\mathrm{where}
\quad 
y_{\max} = \underset{i \in \{1,\ldots,M\}}{\max} 
\{\yi\}
.
\end{equation}
The proof is given in \cite{chouzenoux:15:aca}.
}

Combining the above derivations, 
we can conclude that
\begin{align}\label{eq:gpg,lips,2}
\mathcal{L}(\nabla \gpg) &=
4C^2 \, \|\A\|_2^2 \, \|\A\|_{\infty}^2 \,
\paren{1 - e^{-\frac{1}{\sigma^2}}}
\, e^{\frac{2 y_{\max} -1}{\sigma^2}
}
\nonumber \\
&+ 2\|\A\|_2^2 \,
\Big|1 - C^2 \, \|\A\|_{\infty}^2 \,
 \paren{1 - e^{-\frac{1}{\sigma^2}}}
\, e^{\frac{2 y_{\max} -1}{\sigma^2}
}
\Big|
.
\end{align}
}
\subsection{Reconstruction results of WF-Gaussian}
\fref{fig:gau,fail} shows the reconstruction results from WF-Gaussian. We tried both line search method and the Fisher information for computing the step size, but neither resulted in a successful recovery.
\begin{figure}[hbt!]
    \centering
    \includegraphics[width=0.6\linewidth]{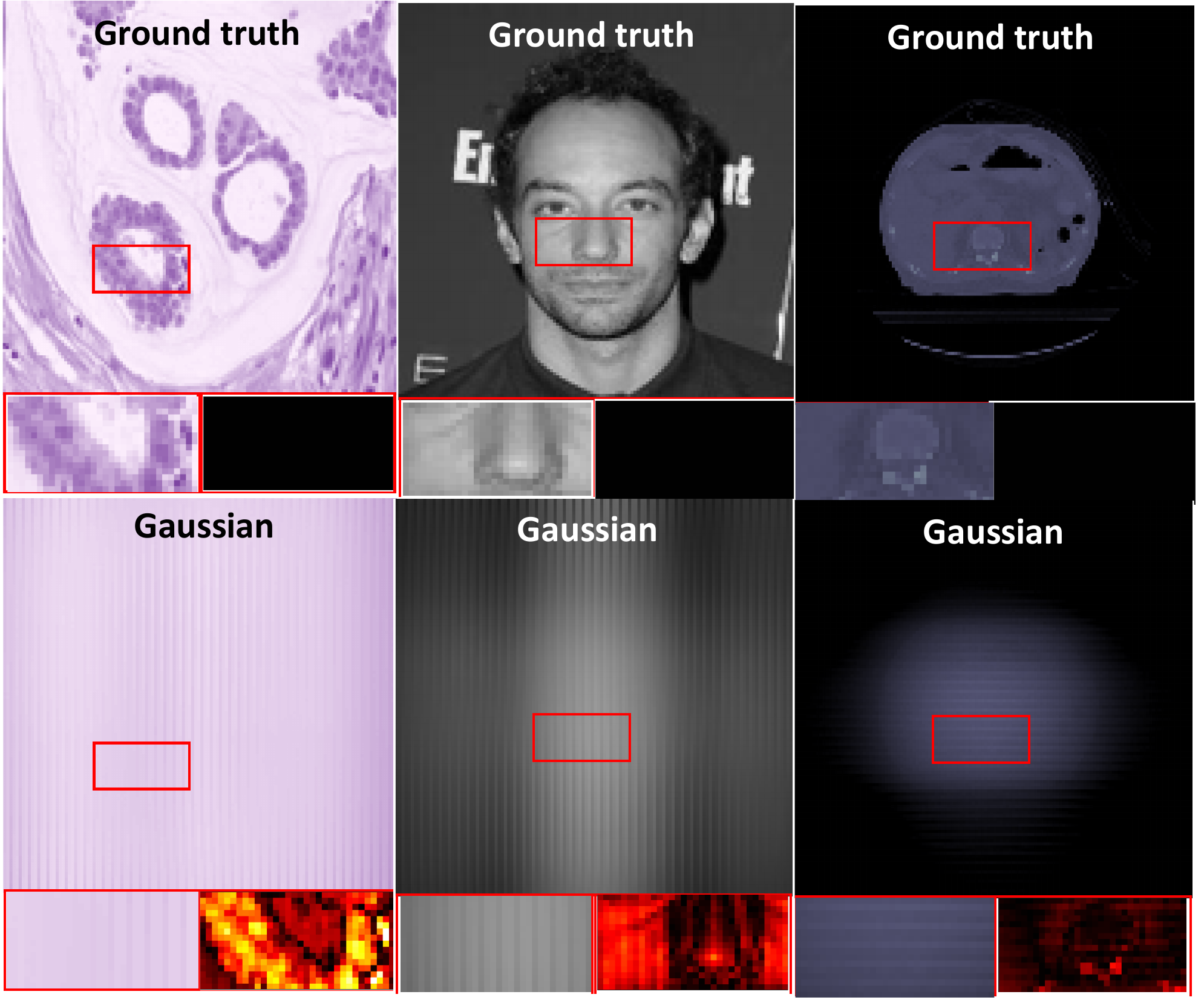}
    \caption{Reconstructed images using WF-Gaussian.}
    \label{fig:gau,fail}
\end{figure}
\subsection{Proof of Theorem 2}
\label{appendix, p2}
It was already shown that
$\nabla \df_{\mathrm{PG}}$
is Lipschitz continuous,
so the remaining problem is to
find a Lipschitz constant for $\s(\x, \sigma)$.
We assume that the data allows the neural network to learn the score function well,
\ie, $ p_\sigma(\x) = p(\x) \circledast \mathcal{N}(0, \sigma^2)$, 
where $\circledast$ denotes (circular) convolution.
We start with some well-known lemmas 
where the proofs can be readily found in \cite{grafakos:04:cam, luenberger:97:obv}. 
We include our proofs just for completeness.

$\textbf{Lemma 2.1:}$
The Fourier transform (and inverse transform) of an absolutely integrable function is continuous.
\\ 
$\textbf{Proof of lemma 2.1:}$ Let $f$ be absolutely integrable and let $\tilde{f}$ be its Fourier transform.
We have 
\begin{multline}\label{eq:proof,lemma2.1}
    |\tilde{f}(w+h) - \tilde{f}(w)| = \left |\int f(x) (e^{-2\pi jx (w+h)} - e^{-2\pi j wx})dx \right |
    \le 
    \int |f(x)| |e^{2\pi jxh}-1|dx \\ 
    \le \max(|e^{2\pi jxh}-1|) \int |f(x)|dx \le 2\int |f(x)|dx.
\end{multline}
Using absolute integrability of $f$, we see 
$|\tilde{f}(w+h) - \tilde{f}(w)|$
tends to 0 as $h$ tends to 0, so $\tilde{f}$ is uniformly continuous, which also implies it is continuous.

The proof of the inverse transform follows similarly.

$\textbf{Lemma 2.2:}$
Suppose a sequence of functions
$f_i: \mathbb{R} \rightarrow \mathbb{R}$
converges in the $L^1$ to some function $f$, 
and that each $f_i$ is absolutely integrable.
Then $f$ is also absolutely integrable.
\\
$\textbf{Proof:}$
Because
\begin{equation}\label{eq:converge,l1}
    \lim_{i \rightarrow \infty} \int_{-\infty}^\infty |f_i(x)-f(x)|dx =0
\end{equation} 
and that $ \int_{-\infty}^\infty |f_i(x)|dx < \infty$. It follows that
\begin{equation}\label{eq:converge,triangle}
     \int_{-\infty}^\infty |f(x)| dx < \int_{-\infty}^\infty |f(x) - f_i(x)| dx  + \int_{-\infty}^\infty |f_i(x)| dx,
\end{equation}
for any $i$.
The second integral is always finite, and for sufficiently large $i$,
the first integral must be finite as it converges to 0.
Hence it is possible to find $i$ such that both integrals converge, so $f$ is absolutely integrable. \qed

\textbf{Proposition 2.1:}
The derivative of 
$\log (p(x) \circledast \mathcal{N}(0, \sigma^2))$
is bounded on the interval $[-C,C]$.
\\
$\textbf{Proof:}$
We start by dropping constant factors and using derivative of a convolution, 
we have 
\begin{equation}\label{eq:derivative, px, app}
    \frac{d}{dx}(\log (p(x) \circledast \mathcal{N}(0, \sigma^2)))
    \sim \frac{\mathcal{F}^{-1}(\imath x \mathcal{F}(p(x)) \cdot
    \mathcal{F}(\mathcal{N}(0, \sigma^2)))}{p(x) \circledast \mathcal{N}(0, \sigma^2)}
    \sim \frac{\mathcal{F}^{-1}(xe^{-x^2} \cdot \mathcal{F}(p(x)))}{p(x) \circledast \mathcal{N}(0, \sigma^2)}
\end{equation}
where $\mathcal{F}$ denotes Fourier transform.
The denominator is continuous and since $x$ lies in a closed interval by assumption,
has a lower bound $M>0$ by the extreme value theorem.
We next consider the numerator.
\\
By \cite[pp.~65]{goodfellow:16:dl}, 
a sequence of Gaussian mixture models (GMMs) can be used
to approximate any smooth probability distribution in $L^2$ convergence. 
Furthermore, $L^2$ convergence implies $L^1$ convergence. 
Hence, consider a sequence of GMMs $f_i$ that converge in 
$L^1$ to $p(x)$. 
By linearity of Fourier transform,
$\mathcal{F}(f_i(x))$ must be a linear combination of terms of the form $e^{-(x-\mu_i)^2/c_i}$ for some $c_i$.
Thus, the numerator $xe^{-x^2} \cdot \mathcal{F}(f_i(x))$ is a finite linear combination
of terms of the form $xe^{-(x-\mu_i)^2/c_i}$, each of which are absolutely integrable.
Therefore, we have a sequence of functions, each of which are absolutely integrable,
that converge in $L^1$ to $xe^{-x^2} \cdot \mathcal{F}(p(x))$,
so by Lemma 2.2, 
this is also absolutely integrable. 
By Lemma 2.1, 
the inverse Fourier transform of this is continuous.
Finally, again by the extreme value theorem and using the boundedness of $x$,
the numerator is bounded above by some $M'>0$.
Hence, the entire expression 
\eqref{eq:derivative, px}
is bounded above by $M'/M$. 

\textbf{Lemma 2.4}:
Suppose we have an everywhere twice differentiable function of two variables
$f(x,y):\reals^2 \rightarrow \reals$. 
Then $\frac{\partial^2}{\partial x \partial y} \log f(x,y)$ is bounded
if the following three conditions are met:
\begin{enumerate}
\item $\frac{\partial^2}{\partial x \partial y} f(x,y)$ is bounded.
\item $f$ itself is bounded below by a positive number and also bounded above.
\item $\nabla f$ is bounded.
\end{enumerate}

\textbf{Proof}:
Suppose we have $f(x,y)$ satisfying those three conditions.
We compute the second partial derivative of its log: 
\begin{equation}\label{eq:dev,log}
    \frac{\partial}{\partial x} \log f(x,y) = \frac{\frac{\partial}{\partial x} f(x,y)}{f(x,y)}
.\end{equation}
and 
\begin{equation}\label{eq:second,dev,log}
    \frac{\partial^2}{\partial x \partial y} \log f(x,y)
    = \frac{(\frac{\partial^2}{\partial x \partial y} f(x,y)) f(x,y) - (\frac{\partial}
    {\partial x} f(x,y))(\frac{\partial}{\partial y} f(x,y))}{f(x,y)^2}
.\end{equation}
From the second condition, the denominator is bounded below by a positive number,
so it suffices to consider the boundedness of the numerator.
The first term of the numerator is a product of two quantities,
the first of which is bounded by the first condition and the second of which is bounded by the second condition.
The second term of the numerator is also a product of two quantities, both of which are bounded by the third condition.
Thus, this shows $\frac{\partial^2}{\partial x \partial y} \log f(x,y)$ is bounded. \qed

\textbf{Proposition 2.2:}
The gradient of $p_\sigma(\x)$ is Lipschitz continuous
on $[-C,C]^N$.
\\
\textbf{Proof:}
By the definition of Lipschitz continuity, 
it suffices to show the Hessian of
$ p_\sigma(\x) = p(\x) \circledast \mathcal{N}(0, \sigma^2 \I)$ has bounded entries.
By renaming the variables, and redefining $f(x,y) = p(x,y, \cdots)$,
we may consider the boundedness of
$\frac{\partial^2}{\partial x \partial y} \left( \log f(x,y) \circledast \mathcal{N}(0, \sigma^2 \I) \right)$
on $[-C,C]^2$.
To apply Lemma 2.4 to remove the log, 
we need to verify the three conditions. 
Define $g(x,y) = f(x,y) \circledast \mathcal{N}(0, \sigma^2 \I)$.
The second condition is readily verified to be true:
By assumption, $x$ and $y$ take values on a closed interval,
thus by the extreme value theorem, so does $g(x,y)$.
Further, $g$ is a convolution of positive numbers and so the output is always positive,
hence, the lower bound of this closed interval is a positive number, verifying this condition. 

For the third condition,
we need to consider boundedness of $\frac{\partial}{\partial x} f(x,y) \circledast \mathcal{N}(0, \sigma^2 \I) $.
This is nearly identical to Proposition 2.2,
with the only difference being we have some general function in terms of only $x$ $f(x,y)$
instead of a probability distribution $p(x)$.
The proof of that lemma is readily adapted for this case
with the only condition needing verification being the absolute integrability over $x$ of $f(x,y)$.
In fact, this is clear because $f$ is always positive;
hence, integrating over $f$ with respect to $x$ must yield a finite number as integrating a second time over $y$ yields 1. 

It thus suffices to consider boundedness of
$h(x,y) = \frac{\partial^2}{\partial x \partial y}
\left( f(x,y) \circledast \mathcal{N}(0, \sigma^2 \I) \right)$.
It is assumed that $f$ is smooth and the convolution of smooth functions is smooth,
which implies $f(x,y) \circledast \mathcal{N}(0, \sigma^2 \I) $ is smooth.
Hence $h$ is differentiable, so it is continuous.
Once again by the EVT, as $x$ and $y$ take values on a closed interval,
$h$ must be bounded.
By Lemma 2.4, the entries of the Hessian of the score function are bounded.
Therefore, a Lipschitz constant of $s_{\btheta} (\x)$ exists. \qed

\textbf{Proof of Theorem 2:}
By Proposition 2.2,
and from the design of Algorithm~\ref{alg:pg_score},
$\x_{t,k}$ and $\bw_{t,k}$ are 
both bounded between 
$[0, C]$ for all $t,k$,
so the Lipschitz constant $\mathcal{L^*}$ 
of $\nabla \df_{\mathrm{PG}}(\cdot)
+
s_{\btheta} (\cdot)$ exists.
With the stepsize $\mu$ satisfying 
$0<\mu<\frac{1}{\mathcal{L^*}}$,
and the weighting factor $\bgamma \in \{0,1\}$
being chosen according to 
whichever higher posterior probability 
between $p(\z|\y,\A,\br)$ and $p(\v|\y,\A,\br)$
(see \cite{li:15:apg}),
then we satisfy all conditions in Theorem 1 of \cite{li:15:apg},
which establishes the critical-point convergence of Algorithm~\ref{alg:pg_score}.
Similar convergence analysis can be found in
\cite{ghadimi:16:agm}. \qed

\subsection{Algorithm Implementation}
\label{appendix, alg}
\textbf{Wirtinger Flow.}
WF is a popular algorithm for phase retrieval.
It first computes the Wirtinger gradient 
(an ascending direction) and then applies gradient descent.
Perhaps the most critical step is to find an appropriate step size.
In this work, we used backtracking line search for Gaussian WF,
and the observed Fisher information \cite{li:22:ppr} for Poisson WF and Poisson-Gaussian WF.
\comment{
Due to the infinite sum in Poisson-Gaussian WF,
we approximate the $s(a, b)$ in \eqref{eq:pois,gau,grad} 
according to \cite{chouzenoux:15:aca}:
\begin{equation}
\label{eq:truncate,sum}
s(a, b)
\approx 
\sum_{n=0}^{n^{+}}
\frac{a^n}{n!}e^{-\paren{\frac{b-n}{\sqrt{2}\sigma}}^2},
\quad 
n^{+} = \lceil n^* + \delta \sigma \rceil,
\end{equation}
with $n^*$ given by 
\begin{align}
\label{eq:lambert}
n^* = \sigma \mathcal{W}\paren{\frac{a}{\sigma^2}e^{b/\sigma^2}}
&\approx
\sigma \paren{
\frac{b}{\sigma^2}
\log
\paren{
\frac{a}{\sigma^2}
}
- \log\paren{
\frac{b}{\sigma^2}
\log
\paren{
\frac{a}{\sigma^2}
}
}
}
\nonumber\\&
= 
\frac{b}{\sigma}
\log
\paren{
\frac{a}{\sigma^2}
}
- \sigma 
\log\paren{
\frac{b}{\sigma^2}
\log
\paren{
\frac{a}{\sigma^2}
}}
,
\end{align}
where $\mathcal{W}(\cdot)$ denotes the Lambert function.
The accuracy of this approximation 
is controlled by $\delta$.
A comprehensive analysis 
on the maximum error value can be found in \cite{chouzenoux:15:aca}.
To further accelerate the computing 
(and avoid floating point overflow) of $\nabla \gpg$,
we observed it effective to use
the gradient of Poisson PR 
for large $\yi$, \eg, $\yi \ge 100$. }
We used this ``trick" in our experiments;
additionally, one can also use 
``defocus" to deal with large $\yi$
\cite{ye:22:set}.
By replacing $\nabla h(\x)$ in Algorithm~\ref{alg:wf}
with the trained score function $s_{\btheta}$,
one can derive the vanilla gradient-descent version
(Algorithm~\ref{alg:pg_score,wfsd})
of the AWFSD algorithm.

\begin{algorithm}[t]
\caption{Poisson-Gaussian phase retrieval via 
WFSD.}
\label{alg:pg_score,wfsd}
\begin{algorithmic}
\Require Measurement \y, system matrix \A, 
step size factor $\epsilon$,
truncation operator $\mathcal{P}_C(\cdot) \rightarrow [0, C]$;
initial image $\x_0$, 
initialize $\sigma_1 > \sigma_2 > \cdots > \sigma_K$.
\For{$k=1:K$}
\For{$t=1:T$}
\State Set step size $\mu = \epsilon \sigma_k^2$.
\State Compute $s_{\btheta} (\x_{t,k}, \sigma_k)$.
\State Set $\x_{t+1,k}=
\mathcal{P}_C\paren{\x_{t,k} - \mu 
\paren{\nabla \gpg(\x_{t,k}) + s_{\btheta} (\x_{t,k}, \sigma_k)}}$.
\EndFor
\EndFor
\\ 
Return $\x_{T,K}$.
\end{algorithmic}
\end{algorithm}

\textbf{PnP-ADMM.}
The plug-and-play ADMM first derives a Lagrangian 
using variable splitting and then applies 
alternating minimization \cite{chan:17:pap}.
In this work, let $\u = \x$, 
and the Lagrangian is
\begin{equation}  
\label{eq:pnp,admm}
L(\x, \u, \bleta; \rho)
= \gpg(|\A\u|^2 + \b) + R(\x)
+ \frac{\rho}{2}
\paren{\| \x - \u + \bleta\|_2^2 - \|\bleta\|_2^2}
.
\end{equation}

\begin{figure*}[ht!]
    \centering
    \includegraphics[width=\linewidth]{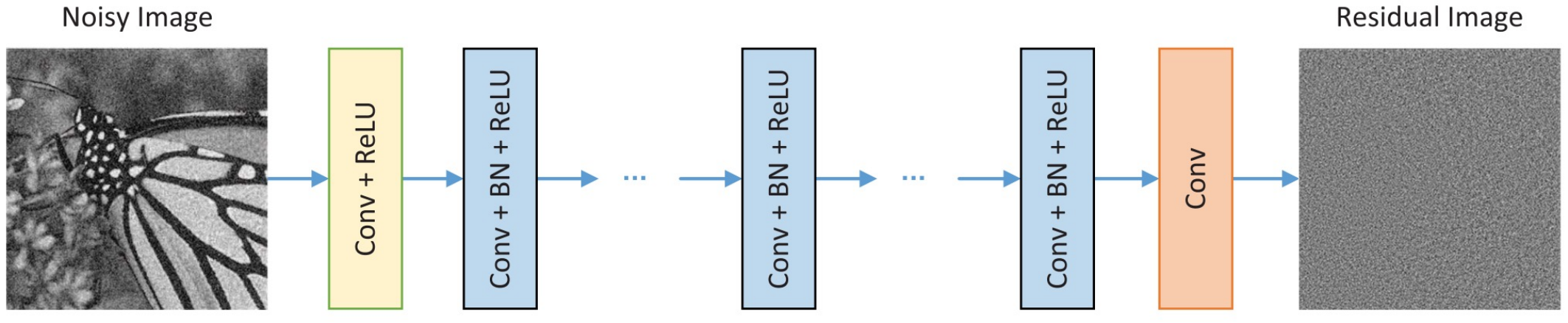}
    \caption{~\emph{The architecture of the adopted DnCNN network~\cite{zhang:17:bag}. }}
    \label{fig:DnCNN}
\end{figure*}

Algorithm~\ref{alg:pnpadmm} summarizes the PnP-ADMM algorithm for phase retrieval.
In this work, we trained the denoiser $h_{\btheta}$ using the network DnCNN~\cite{zhang:17:bag}.
As shown in Fig.~\ref{fig:DnCNN},
the architecture of DnCNN consists of convolution (Conv), Rectified Linear Unit (ReLU), and batch normalization(BN).
The network was trained with residual learning
where the output is the noise residual and the clean image was obtained by subtracting the noisy output.
We trained all denoisers for 400 epochs with image patches of size $40 \times 40$ on each given dataset. 
\begin{algorithm}[ht]
\caption{Poisson-Gaussian phase retrieval via PnP-ADMM}
\label{alg:pnpadmm}
\begin{algorithmic}
\Require Measurement $\y$, system matrix $\A$, 
initialization of image $\x_0$, 
initialization of auxilary variable $\u_0 = \x_0$,
initialization of dual variable $\bleta_0 = \blmath{0}$,
pre-trained denoiser $h_{\btheta}$,
Lagrangian penalty parameter $\rho$.
\For{$k=1:K$}
\For{$t=1:T$}
\State Compute step size $\mu_{t,k}$.
\State Set 
$\u_{t+1,k} = \u_{t,k} - \mu_{t,k} 
\paren{\nabla \gpg(\u_{k,t})
+\rho (\u_{t,k} - \xk - \bleta_{k})}
$.
\EndFor
\State Set $\xkk = h_{\btheta}(\xk)$.
\State Set $\bleta_{k+1} = \bleta_{k} + \xkk - \ukk$.
\EndFor
\end{algorithmic}
Return $\x_{K}$.
\end{algorithm}

\textbf{PnP-PGM.}
Similar to PnP-ADMM, one can also derive a proximal gradient method as shown in Algorithm~\ref{alg:pnppgm}.
Here we assume the denoising of $h_{\btheta}$ 
is a proximal operation.
\comment{
One might be skeptical about the weighted average 
at the first glimpse,
but it turns out that it leads to a denoiser
with a regularization parameter
that depends on $\beta$ and $\lambda$.
For example, if $h_{\btheta}(\u) = \frac{1}{1+\lambda}\u$, then Algorithm~\ref{alg:pnppgm}
is performing gradient descent:
\begin{equation}\label{eq:pnp,prox,gd}
\xkk = \xk - (1-\beta)\mu_k
\paren{\nabla \gpg (\xk) + \frac{\lambda \beta}{\mu_k(1-\beta)(1+\lambda)}\xk},
\end{equation}
which corresponds to ridge regression
with a particular \red{iteration-dependent} regularization parameter.
}

\begin{algorithm}[ht]
\caption{Poisson-Gaussian phase retrieval via PnP-PGM}
\label{alg:pnppgm}
\begin{algorithmic}
\Require Measurement $\y$, 
system matrix $\A$, 
initialization of image $\x_0$, 
pre-trained denoiser $h_{\btheta}$,
averaging factor $\beta$.
\For{$k=1:K$}
\State Compute step size $\mu_{k}$.
\State Set 
$\tilde{\x}_k = \xk - \mu_k \nabla \gpg(\xk)$.
\State 
Set $\bar{\x}_k = h_{\btheta}(\tilde{\x}_k)$.
\State Set $\xkk = (1 - \beta)\tilde{\x}_k + 
\beta \bar{\x}_k$.
\EndFor
\end{algorithmic}
Return $\x_{K}$.
\end{algorithm}

\textbf{SD-RED.}
Regularization by denoising (RED) is an alternative
to PnP methods that is based on an
explicit image-adaptive regularization functional:
$\frac{1}{2}\x'\paren{\x - h_{\btheta}(\x)}$.
This regularizer reflects
the cross-correlation between the image and its denoising residual \cite{romano:17:tle}.

Algorithm~\ref{alg:red} summarizes the RED approach
for phase retrieval.
\begin{algorithm}[ht]
\caption{Poisson-Gaussian phase retrieval via RED}
\label{alg:red}
\begin{algorithmic}
\Require Measurement $\y$, 
system matrix $\A$, 
initialization of image $\x_0$, 
pre-trained denoiser $h_{\btheta}$,
regularization factor $\beta$.
\For{$k=1:K$}
\State Compute stepsize $\mu_{k}$.
\State Set 
$\xkk = \xk - \mu_k \paren{\nabla \gpg (\xk) 
+ \beta (\xk - h_{\btheta}(\xk))}$.
\EndFor
\end{algorithmic}
Return $\x_{K}$.
\end{algorithm}

\textbf{SD-RED-SELF.}
Other than supervised denoising approaches,
we also implemented a self-supervised denoising method
known as ``noise2self" \cite{batson:19:nbd}, 
which designed a neural network
to be $\mathcal{J}$-invariant so that
the self-supervised loss can be represented as
the sum of supervised loss and the variance of noise.
The ``SD-RED-SELF" algorithm
refers to training $h_{\btheta}$ in Algorithm~\ref{alg:red}
in such self-supervised fashion on each test data.

\begin{figure}[ht]
    \centering
    \includegraphics[width=\linewidth]{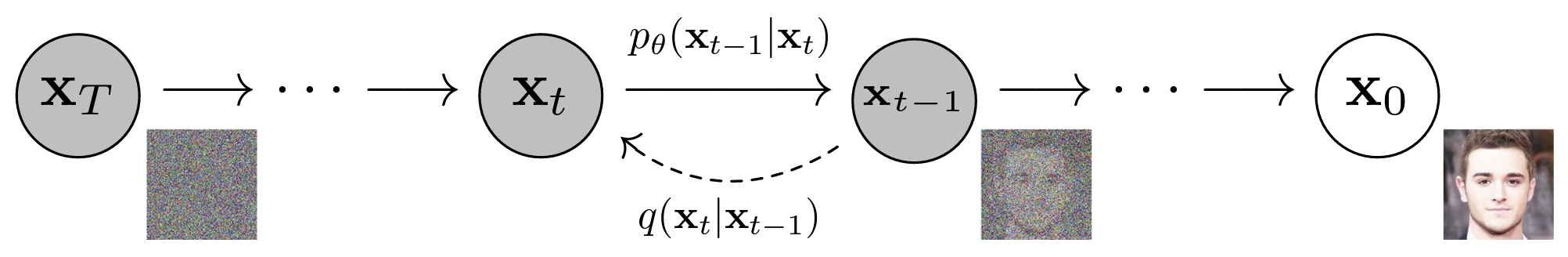}
    \caption{~\emph{The DDPM framework,
    adopted from \cite{ho:20:ddp}.}}
    \label{fig:ddpm,framework}
\end{figure}

\textbf{DOLPH.}
DOLPH is based on the DDPM model \cite{ho:20:ddp},
which first gradually adds Gaussian noise to data according to a variance schedule $\beta_1, \cdots, \beta_T$
so that $q(\x_t | \x_{t-1}) = \mathcal{N}(\x_t; \sqrt{1-\beta_t} \x_{t-1}, \beta_t \I)$,
as illustrated in \fref{fig:ddpm,framework}. 
Using the notation $\alpha_t = 1-\beta_t$ and $\overline{\alpha_t} = \prod_{s=1}^t \alpha_s$, we have 
\begin{equation}
    q(\x_t | \x_0) = \mathcal{N}(\x_t; \sqrt{\overline{\alpha_t}} \x_0, (1-\overline{\alpha_t}) \I).
\end{equation}
It can be shown \cite{ho:20:ddp} that the appropriate loss function to use is 
\begin{equation}
    L(\theta) = \mathbb{E}_{t, \x_0, \epsilon} \left [ \| \epsilon - \epsilon_\theta(\sqrt{\overline{\alpha_t}} \x_0 + \sqrt{1-\overline{\alpha_t}} \epsilon, t)\|^2\right],
\end{equation}
where $\epsilon$ is selected from $\mathcal{N}(0, \I)$.
The sampling and reconstruction algorithm requires the addition of noise each step as shown in the following algorithm.
Experimentally, however, we found that setting $\sigma_t=0$ results in higher quality reconstructed images.
Furthermore, we choose $T=100$, $\beta_1=10^{-4}$, and $\beta_T=0.3$.
For the theory to hold, $\x_T$ should be indistinguishable from white Gaussian noise,
which is readily verified to be true for these parameters.
Finally, the stepsize $\mu_k$ of the gradient descent step can be chosen
according to the Lipschitz constant of the Poisson-Gaussian likelihood to ensure convergence,
or empirically, as is done in the experiments.

\section*{Acknowledgement}
The authors appreciate thoughtful discussions 
with Arian Eamaz and Farhang Yeganegi
from the University of Illinois Chicago (UIC). 

\end{document}